\documentclass[usenatbib,useAMS]{mnras}

\usepackage{graphicx}	
\usepackage{amsmath}	
\usepackage{amssymb}	
\usepackage{multicol}        
\usepackage{bm}		
\usepackage{pdflscape}	

\def \aap{A\&A}
\def \apjl{ApJ}

\def \mnras{MNRAS}
\def \apj{ApJ}
\def \aaps{A\&AS}
\def \aj{AJ}

\def \apjs{ApJS}
\def \nat{Nature}

\def \araa{ARA\&A}
\bibpunct{(}{)}{;}{a}{}{,}

\defcitealias{buder20}{B20}
\defcitealias{zwitter18}{Z18}
\defcitealias{brown20}{Gaia eDR3}
\defcitealias{carrera19}{C19}

\newcommand{\teff}{T_\mathrm{eff}}
\newcommand{\logg}{\log g}
\newcommand{\feh}{\mathrm{[Fe/H]}}
\newcommand{\alphafe}{\mathrm{[\alpha/Fe]}}

\newcommand{\kms}{km~s$^{-1}$}

\usepackage[T1]{fontenc}
\usepackage{ae,aecompl}

\usepackage{newtxtext,newtxmath}
\usepackage{CJK}

\begin{document}
\begin{CJK*}{UTF8}{gbsn}
\label{firstpage}
\pagerange{\pageref{firstpage}--\pageref{lastpage}}


\title[The GALAH$+$ Survey: Radial Velocities]{The GALAH$+$ Survey: A New Library of Observed Stellar Spectra Improves Radial Velocities and Hints at Motions within M67
}

\author[T. Zwitter et al.]{Toma\v{z} Zwitter,$^{1}$\thanks{Contact e-mail: \href{mailto:tomaz.zwitter@fmf.uni-lj.si}{tomaz.zwitter@fmf.uni-lj.si}}
Janez Kos,$^{1}$ Sven~Buder,$^{2,3}$ Klemen \v{C}otar,$^{1}$ 
Martin Asplund,$^{4}$\newauthor
Joss~Bland-Hawthorn,$^{5,3}$
Andrew~R.~Casey,$^{6,7}$
Gayandhi~M.~De~Silva,$^{8,9}$\newauthor
Valentina~{D'Orazi},$^{10}$
Ken~C.~Freeman,$^{2}$
Michael~R.~Hayden,$^{5,3}$
Geraint~F.~Lewis,$^{5}$\newauthor
Jane~Lin,$^{2,3}$
Karin~Lind,$^{11,12}$
Sarah~L.~Martell,$^{13,3}$
Katharine~J.~Schlesinger,$^{2}$\newauthor
Sanjib~Sharma,$^{5,3}$
Jeffrey~D.~Simpson,$^{13,3}$
Dennis Stello,$^{5,3,13,14}$
Daniel~B.~Zucker,$^{15,9}$\newauthor
Kevin~L.~Beeson,$^{1}$ 
Richard~de~Grijs,$^{15,16}$
Thomas Nordlander,$^{2,3}$\newauthor
Yuan-Sen~Ting (丁源森),$^{17,18,19,2}$
Gregor Traven,$^{20}$
Rok Vogrin\v{c}i\v{c},$^{1}$
Fred~Watson,$^{21}$\newauthor
Rob Wittenmyer$^{22}$
\\
\\
(Affiliations listed after the references)}

\date{}

\pubyear{2020}

\maketitle
\end{CJK*}

\begin{abstract}
{\sl GALAH$+$} is a magnitude-limited survey of high resolution stellar spectra obtained by the HERMES spectrograph at the Australian Astronomical Observatory. Its third data release provides reduced spectra with new derivations of stellar parameters and abundances of 30 chemical elements for 584,015 dwarfs and giants, 88\%\ of them in the {\sl Gaia} magnitude range $11 < G < 14$. Here we use these improved values of stellar parameters to build a library of observed spectra which is useful to study variations of individual spectral lines with stellar parameters. This and other improvements are used to derive radial velocities with uncertainties which are generally within 0.1~km~s$^{-1}$ or $\sim 25$\%\ smaller than in the previous release. Median differences in radial velocities measured here and by the Gaia DR2 or {\sl APOGEE} DR16 surveys are smaller than 30~m~s$^{-1}$, a larger offset is present only for Gaia  measurements of giant stars. We identify 4483 stars with intrinsically variable velocities and 225 stars for which the velocity stays constant over $\geq 3$ visits spanning more than a year. The combination of radial velocities from {\sl GALAH$+$} with distances and sky plane motions from {\sl Gaia} enables studies of dynamics {\it within} streams and clusters. For example, we { estimate} that the open cluster M67 has a total mass of $\sim 3300$~M$_\odot$ { and its outer parts seem to be expanding, though astrometry with a larger time-span than currently available from Gaia eDR3 is needed to judge if the latter result is real.} 
\end{abstract}




\section{Introduction}

Galactic archaeology \citep{freeman02} aims to decipher the structure and formation of our Galaxy as one of the typical galaxies in the universe through detailed measurements of stellar kinematics and chemistry of their atmospheres. Recent studies of Galactic dynamics show that the disc is not an axisymmetric equilibrium structure, but dynamically young and perturbed, also by the on-going passages of the Sagittarius dwarf galaxy \citep{antoja18,helmi18,blandhawthorn19,helmi20}. Such perturbations inflict variations in stellar positions and velocities which are much smaller than their nominal values. Fortunately, astrometric measurements from the second data release of the {\sl Gaia} mission of the European Space Agency \citep{brown18} provide stellar coordinates, parallaxes and proper motions with an exquisite accuracy never seen before. The radial velocity spectrograph on board the same satellite \citep{cropper18} is reporting also the radial velocity (RV) measurements for an unprecedented number of over 7 million stars \citep{katz19}. While the RV precision for bright stars ($G_{RVS} \in [4, 8]$~mag) is between 0.22 and 0.35 \kms\ it worsens to 1.4 \kms\ for stars with $G_{RVS} = 11.75$~mag and effective temperature of 5000~K.  

{\sl Gaia} eDR3 proper motion and parallax measurements allows us to measure velocities of stars in the plane of the sky at very high accuracy. For example, a solar type or red clump star at a distance of 1~kpc with a velocity of 9~\kms\ (in the plane of the sky) has a Gaia-based uncertainty of only $\sim 0.2$~\kms, and stars moving slower have even smaller uncertainties. Hence, it is desirable to have the perpendicular line-of-sight RVs measured at a similar level of accuracy. \citet{steinmetz20} presents the final data release of the 10-yr {\sl RAVE} survey. It reports RVs of 518,387 spectra with a typical accuracy of 1.4~\kms. The {\sl Gaia-ESO} survey \citep{gilmore12} iDR5 lists a smaller number but fainter targets at similar levels of accuracy. The on-going {\sl LAMOST} medium resolution survey \citep{liu20} aims for comparable uncertainty, but for a much larger number of spectra of brighter stars. {\sl APOGEE} DR16 \citep{jonsson20} includes 473,307 spectra, mostly from the Northern hemisphere, with similar precision as reported in this paper (see below). We note that, contrary to our approach, none of these surveys calculates RV taking into account  convective shifts within the stellar atmosphere and gravitational redshift of light as it travels to the distant observer.

Here we describe the derivation of RVs with uncertainties typically smaller than 0.1~\kms, though for a dozen-times smaller set than derived by {\sl Gaia}. The cornerstone are new values of effective temperature, surface gravity, metallicity and $\alpha$-enhancement for 584,015 spectra from the third data release of the {\sl GALAH$+$} survey \citep[][hereafter B20]{buder20} which presents also an unprecedented set of measurements of abundances of 30 chemical elements ([X/Fe]) for the same stars. \citet{sharma20} and \citet{hayden20} show how stellar ages can be inferred for the same objects. Derivation of accurate RVs builds on a procedure described earlier \citep[][hereafter Z18]{zwitter18}, but better parameter values and a number of procedure improvements now make the uncertainties $\sim 25$~\%\ smaller, and allow RVs to be derived for 72\%\ more spectra. 

The paper is organised as follows: in the next section we briefly discuss the observational data and the reduction pipeline. In Section \ref{Sectemplates} we present a library of median-combined observed spectra across the stellar parameter space. Section \ref{SecRVs} discusses the RV measurement pipeline. Section \ref{Secrepeats} uses repeated observations of the same stars to identify candidates with constant and with variable RVs. Section \ref{SecM67} illustrates the reach of these results, with a discussion of stellar motions within the cluster M67 used as an example. Section \ref{Secconclusions} contains the final remarks and Section \ref{Secdataproducts} discusses the data products.

\section{Observational data and their reductions}

{\sl GALAH+} includes data from ambitious stellar spectroscopic surveys which use the HERMES spectrograph that simultaneously observes up to 392 stars within the $\pi$ square degree field of the 3.9-m Anglo-Australian Telescope at the Australian Astronomical Observatory (AAO) at Siding Spring. The surveys are GALAH Phase 1 \citep[bright, main, and faint survey, 70\% of all data,][]{desilva15}, K2-HERMES \citep[17\%,][]{wittenmyer18,sharma19} and TESS-HERMES \citep[5\% of data,][]{sharma18}, as well as additional  GALAH-related projects \citep[8\%,][]{martell17}, including observations of the bulge and a number of stellar clusters. Spectra cover 4 wavelength ranges: 4713--4903 \AA\ (blue arm), 5648--5873 \AA\ (green arm),  6478--6737 \AA\ (red arm), and 7585--7887 \AA\ (infra-red arm) at a resolving power of $R = 28,000$. The median S/N per pixel in the green arm is $\sim 35$. For 88.8\%\ of the  targets, which are within $G \in [11.0,14.0]$~mag, this is achieved after three consecutive 20-min exposures, others require shorter or longer sequences. Data from a given star from such an uninterrupted sequence is called a spectrum in this paper. Its effective time of observation is assumed to be mid-time of the sequence.

The data reduction pipeline is described in \citep{kos17}. We use results of its version 5.3. From the pipeline products, we use the wavelength calibrated spectra in ADU counts (no normalisation of continuum) with preserved pixel binning (no resampling). Altogether, 694,459 spectra collected between 16 Nov 2013 and 25 Feb 2019 are considered, but with additional requirements on their physical characterisation as reported in the $3^\mathrm{rd}$ {\sl GALAH$+$} data release \citepalias{buder20}. In particular, we require the values of effective temperature ($\teff$), surface gravity ($\logg$), $\feh$ and $\alphafe$ are all available in \citetalias{buder20}. These values are now much more accurate compared with the previous data release \citep{buder18}, which translates into a more consistent definition of the observed stellar templates (Section \ref{Sectemplates}) and smaller uncertanties in the RVs (Section \ref{SecRVs}). Median formal errors are now 98~K in $\teff$, 0.19~dex in $\logg$, 0.088~dex in $\feh$, and 0.045~dex in $\alphafe$. These error estimates are conservative; Table 2 of \citetalias{buder20} reports about a third better accuracy at S/N~$=40$. Our final selection contains 579,653 spectra for which RVs are determined. Note that 117,726 of these spectra have the value of reduction flag\,  {\sl flag\_sp }$>0$, indicating problems with spectral peculiarities, data reduction or spectrum analysis. From these 48,638 spectra have astrometric index RUWE~$>1.4$ \citep[these stars may not be consistent with a single-star astrometric solution,][]{brown18}, 18,058 have raised binarity or emission object flags \citep{traven17,traven20}, 19,131 have a very low S/N ratio ($S/N<10$) and others suffer from various reduction or convergence issues. RVs of spectra with {\sl flag\_sp }$>0$ are reported, but they need to be treated with caution.

\section{Library of observed spectral templates}
\label{Sectemplates}

Similarly to \citetalias{zwitter18}, we calculate RVs in a two-step process: the observed spectra can be noisy, so we first use a large number of spectra with very similar values of stellar parameters to construct a nearly noise-free observed spectral template which is then compared to synthetic spectra. Such an approach yields better results than direct correlation between observed and synthetic spectra and allows control of systematics, as discussed below. The workflow is similar to the one in \citetalias{zwitter18}, so we do not repeat its description here, but only emphasise the differences.

Spectra are grouped according to values of four parameters: $\teff$, $\logg$, $\feh$, and $\alphafe$. The alpha abundance is added here because there are a number of scientific applications where a distinction of observed spectra by $\alpha$ enhancement is important. Also, the parameter bins are now different: following the parameter uncertainties mentioned above their values are rounded to the nearest step in the $N \Delta \teff$ ladder in temperature, $N \Delta \logg$ in gravity, $N \Delta \feh$ in iron abundance, and $N \Delta \alphafe$ in $\alpha$ enhancement, where $N$ is an integer and $\Delta \teff = 200$~K, $\Delta \logg = 0.3$~dex, $\Delta \feh = 0.17$~dex, and $\Delta \alphafe = 0.09$~dex. These rounded values now serve as labels that indicate to which stellar parameter bin our spectrum belongs. Observed spectra are shifted to a common reference frame using RV values from \citetalias{buder20} which are accurate to $\sim 0.4$~\kms, hence better than the Guess values used in \citetalias{zwitter18}.  

A meaningful median spectrum can be derived only if we combine a sufficient number of observed spectra within a given parameter bin. We adopt a threshold of 100 spectra per bin, with the additional requirement that they have {\sl flag\_sp}~$=0$, thus excluding spectra with peculiarities or reduction problems. There are 718 bins with at least 100 {\sl flag\_sp}~$=0$ spectra, so this is also the size of the library of observed spectral templates. The most populous bin is located at the main sequence turn-off ($\teff=6000$~K, $\logg=4.2$, $\feh=0.0$, $\alphafe = 0.0$) and contains 7290 spectra, from these 6315 have {\sl flag\_sp}~=0. 

The spectra are interpolated to a log-spaced grid which is $\sim 3$-times denser than the observed one (12288 points over each of the 4713--4900, 5468--5871, 6478--6736, 7693--7885\AA\ intervals). Next they are normalised with a three-piece cubic spline with symmetric 3.5~$\sigma$ rejection levels and 10 iterations. The combined spectrum is calculated as a weighted median of spectra, with weights proportional to the square of the S/N ratio in the red channel, truncated at S/N~$=200$. This is a better choice than a simple median used in \citetalias{zwitter18}. However, such a weighted median could potentially be driven by a small number of very high S/N spectra in a given parameter bin; we checked that this is not the case. 

\begin{figure}
  \includegraphics[width=0.78\columnwidth,angle=270]{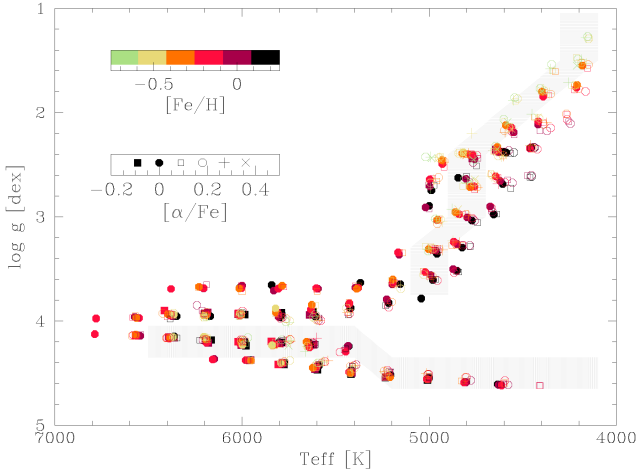}                  
  \caption{Kiel diagram of a library of observed median spectra. Their iron abundances are colour coded and different symbols mark the $\alpha$-enhancement bins, as indicated in the legend. Medians in grey areas along the main sequence and the red giant branch are used for plotting of equivalent widths of spectral lines in Figures \ref{Figmed1}--\ref{Figmed6}.}
  \label{FigKiel}
\end{figure}

Figure \ref{FigKiel} plots the 718 spectral bins in a Kiel diagram. The position of each bin is given as a weighted median of $\teff$ and $\logg$ values of its spectra. So the symbols with colour-coded iron abundances and $\alpha$-enhancements plotted with different symbols do not overlap completely. The figure demonstrates a good coverage of stellar evolutionary tracks, with the exception of hot or very cool stars, which are rarely observed by {\sl GALAH}.

\begin{figure}
  \includegraphics[width=0.79\columnwidth,angle=270]{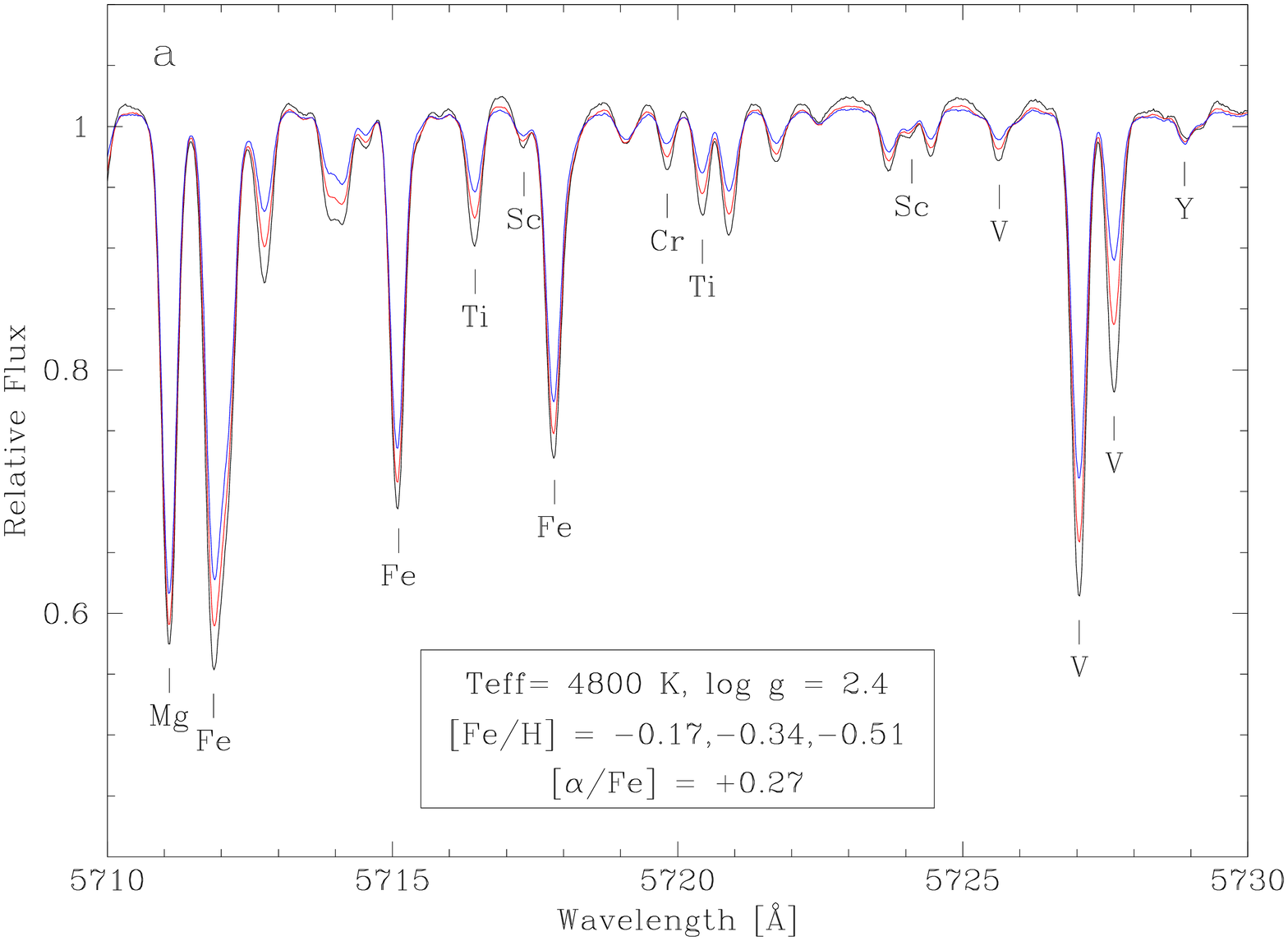} \\
  \includegraphics[width=0.79\columnwidth,angle=270]{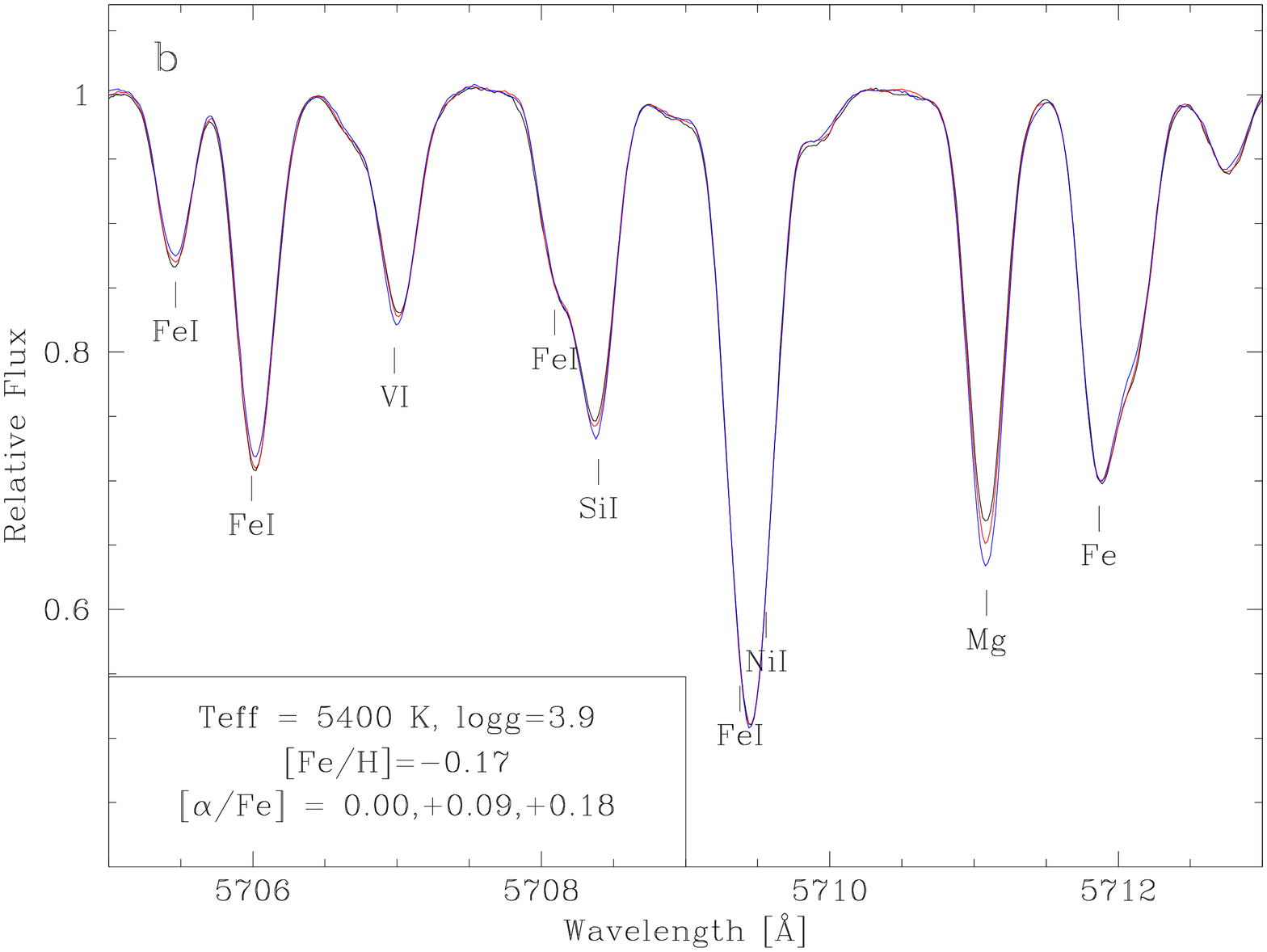}                  
  \caption{A sequence of median observed spectra in the green arm, varying their iron abundance (a) or $\alpha$-enhancement (b).}
  \label{Figexamplemedians}
\end{figure}

Figure \ref{Figexamplemedians} shows two series of observed median spectra if all but one parameter is kept constant. Panel {\sl a} shows how spectra change with $\feh$, and panel {\sl b} demonstrates a variation with $\alphafe$. Understandably, the former affects the depth of all spectral lines, while the latter shows variation mostly in lines of $\alpha$ elements. A moderately different continuum level of spectra in panel {\sl a} is due to normalisation process: we use symmetric rejection criteria so the continuum is at a level a bit larger than $1.0$, depending on the strength of absorption lines. Note that even when [Fe/H] is kept constant (panel {\sl b}), the strengths of the Fe lines vary slightly because of slight differences in mean $\teff$\ and $\feh$\ between the different $\alphafe$ groups.

\begin{figure}
  \includegraphics[width=0.78\columnwidth,angle=270]{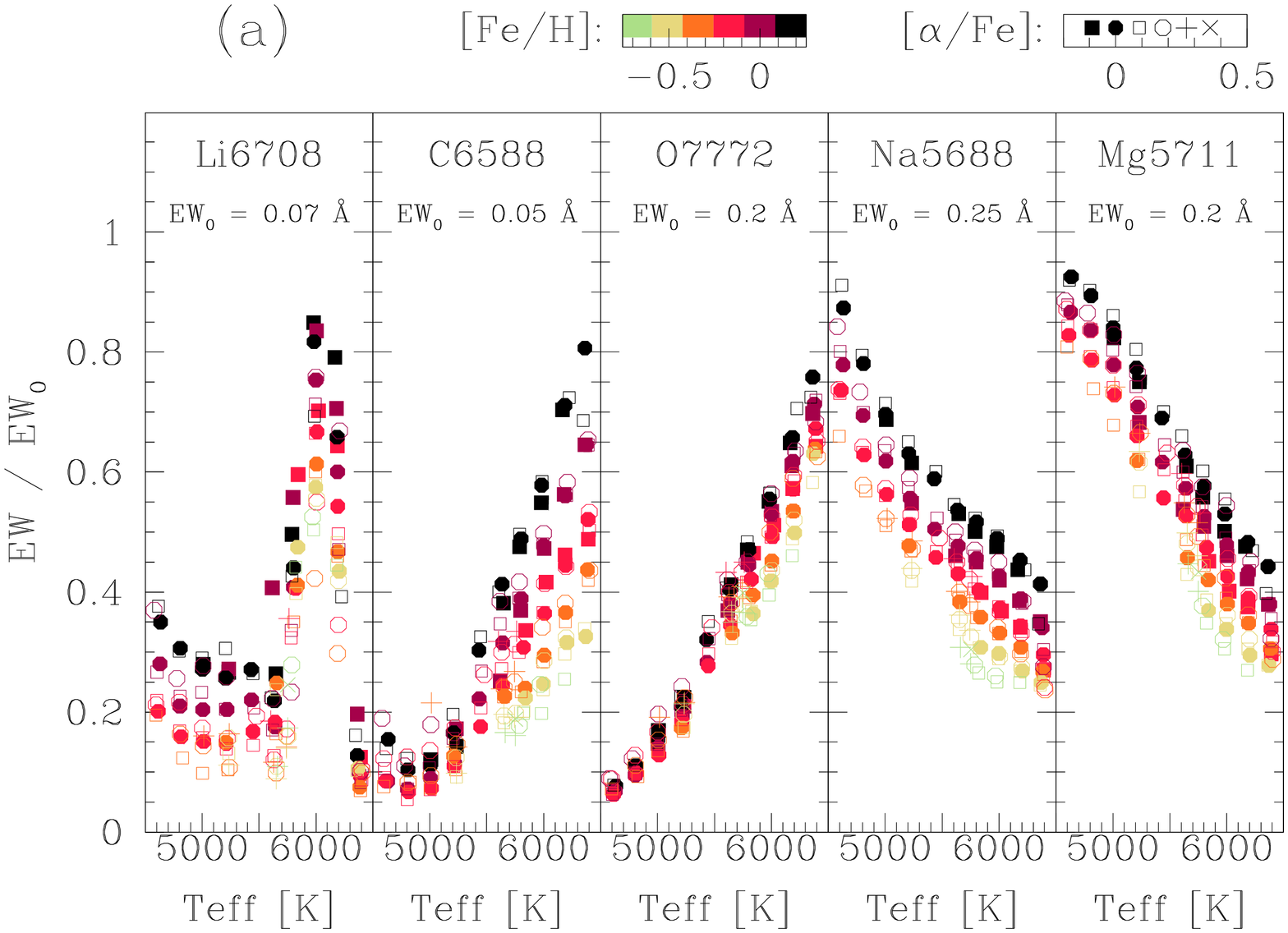} \\                
  \includegraphics[width=0.78\columnwidth,angle=270]{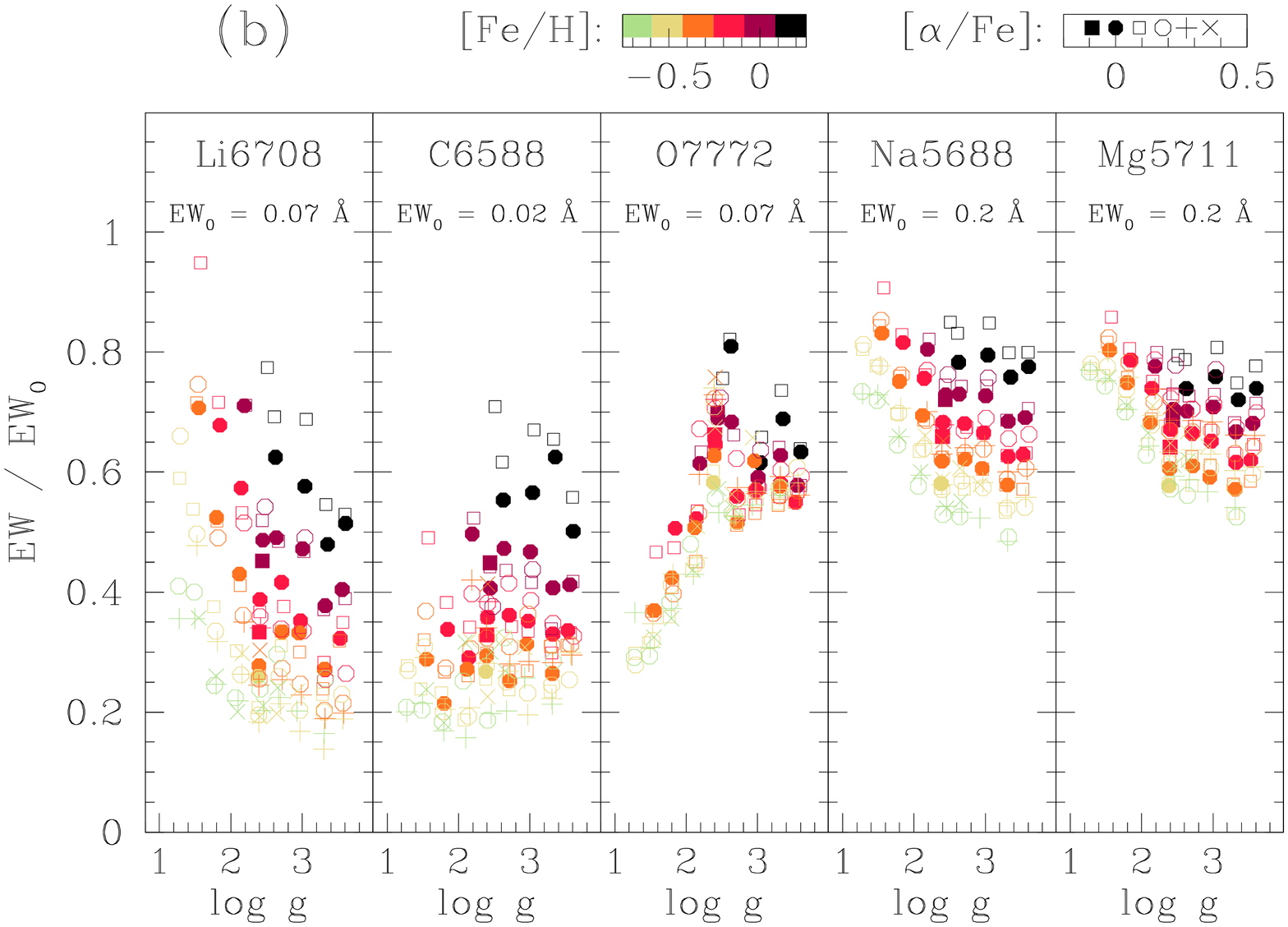}                 
  \caption{Equivalent widths of the strongest lines of elements with $3 \leq Z \leq 12$ along the main sequence (a) and the red giant branch (b). Iron abundances are colour coded, different symbols mark the $\alpha$-enhancement bins, as indicated in the legend.}
  \label{Figmed1}
\end{figure}

\begin{figure}
  \includegraphics[width=0.78\columnwidth,angle=270]{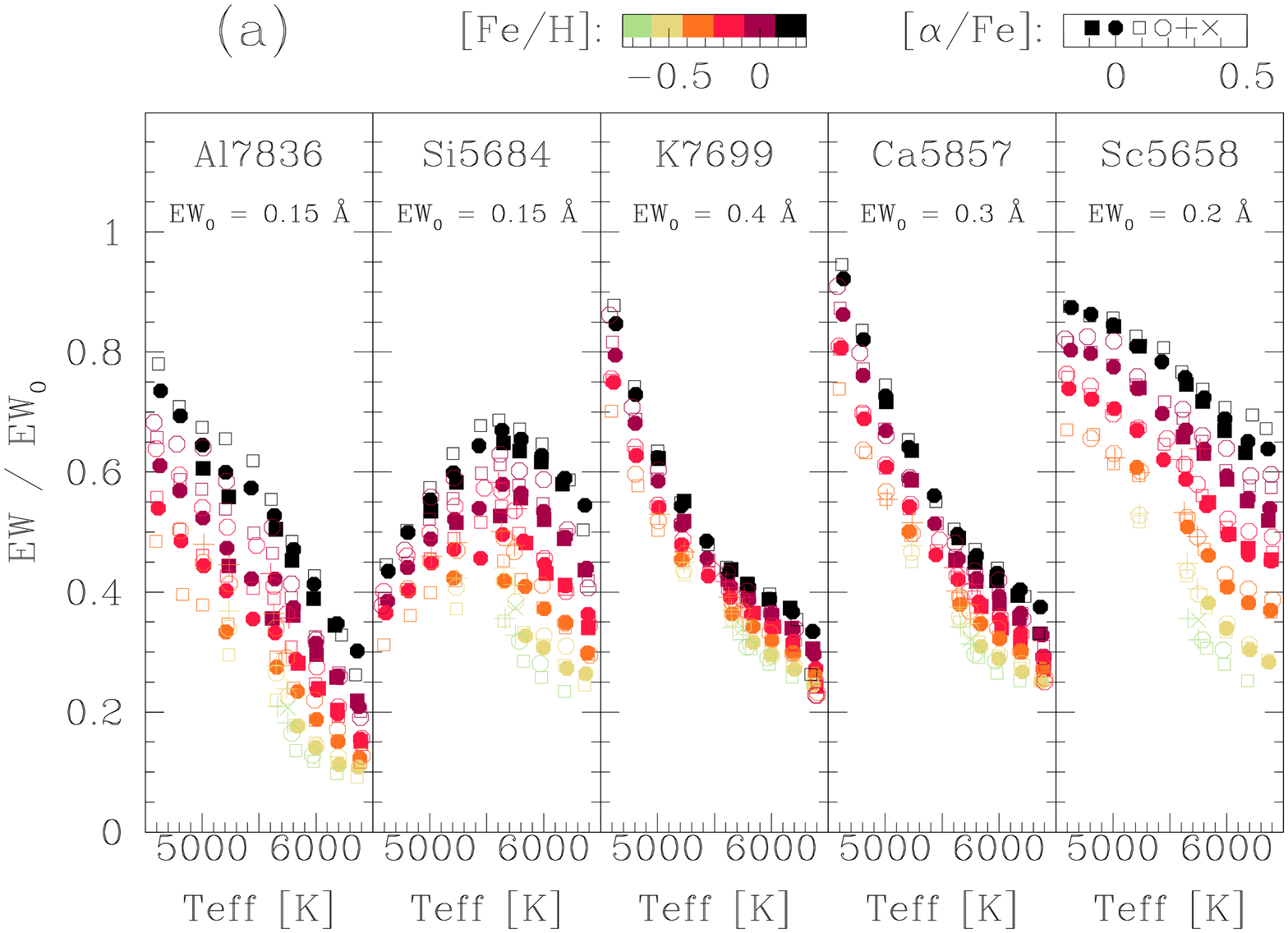} \\                
  \includegraphics[width=0.78\columnwidth,angle=270]{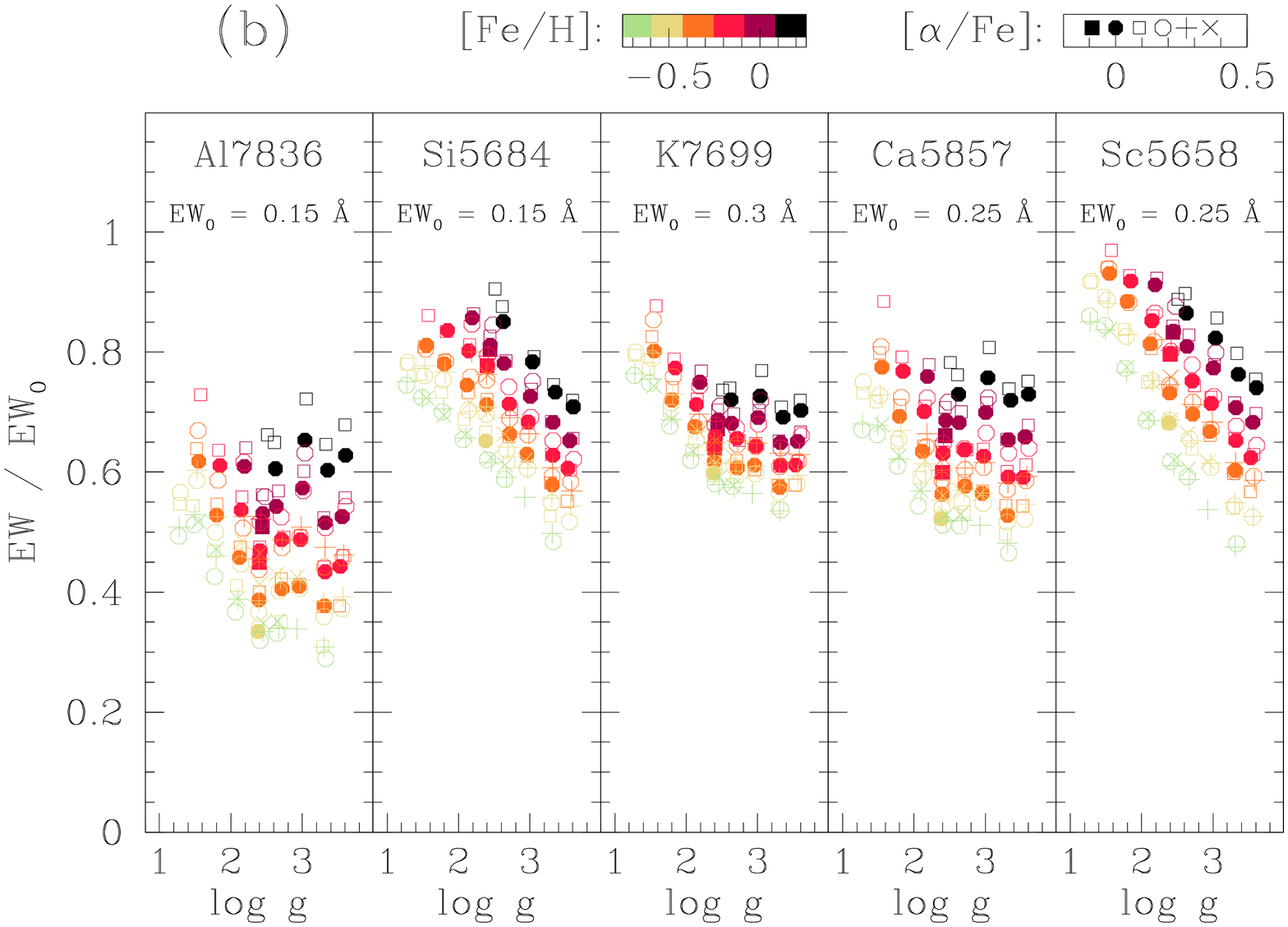}                 
  \caption{As Figure \ref{Figmed1}, but for lines of chemical elements with $13 \leq Z \leq 21$.}
  \label{Figmed2}
\end{figure}

\begin{figure}
  \includegraphics[width=0.78\columnwidth,angle=270]{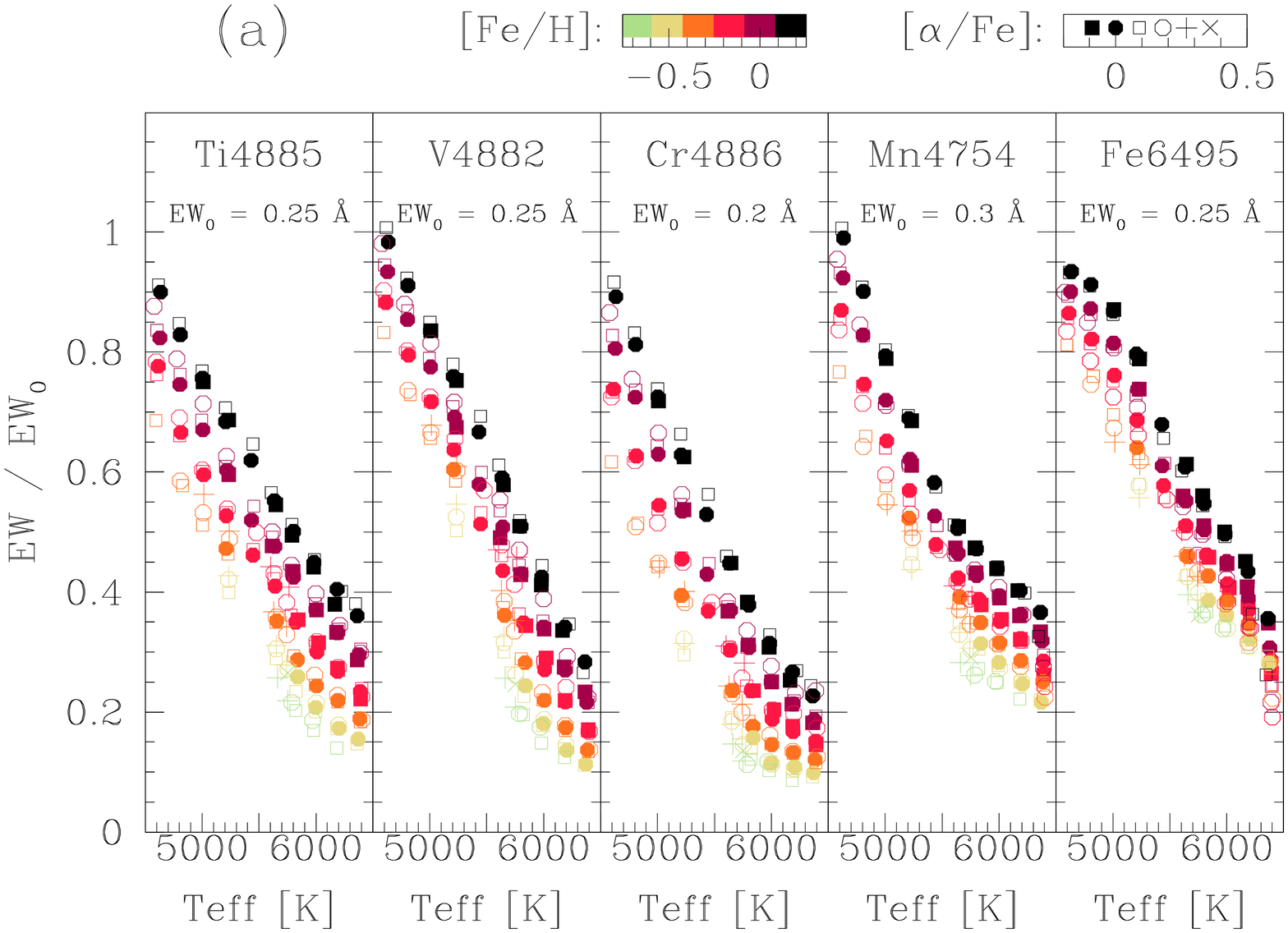} \\                
  \includegraphics[width=0.78\columnwidth,angle=270]{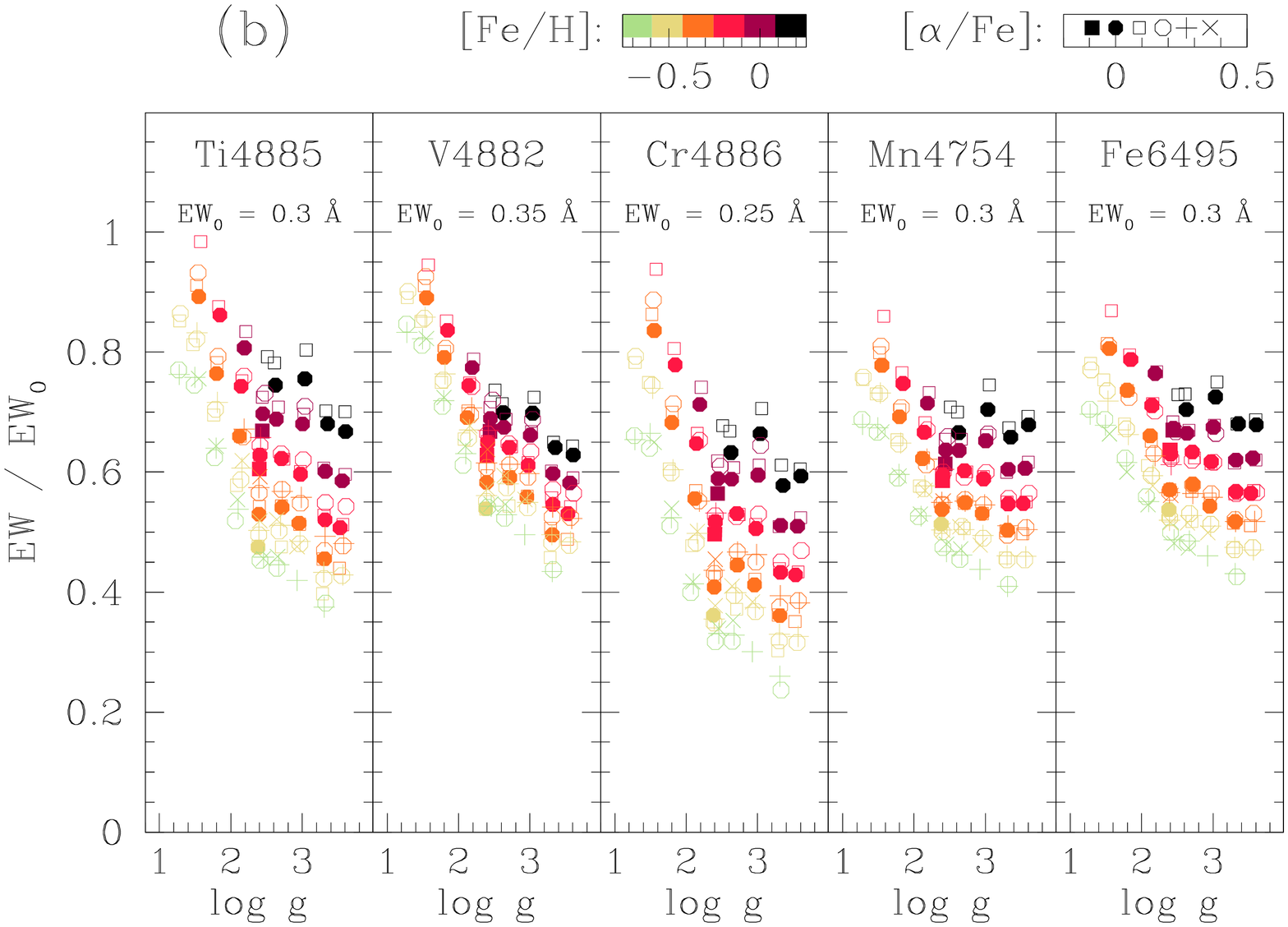}                 
  \caption{As Figure \ref{Figmed1}, but for lines of chemical elements with $22 \leq Z \leq 26$.}
  \label{Figmed3}
\end{figure}

\begin{figure}
  \includegraphics[width=0.78\columnwidth,angle=270]{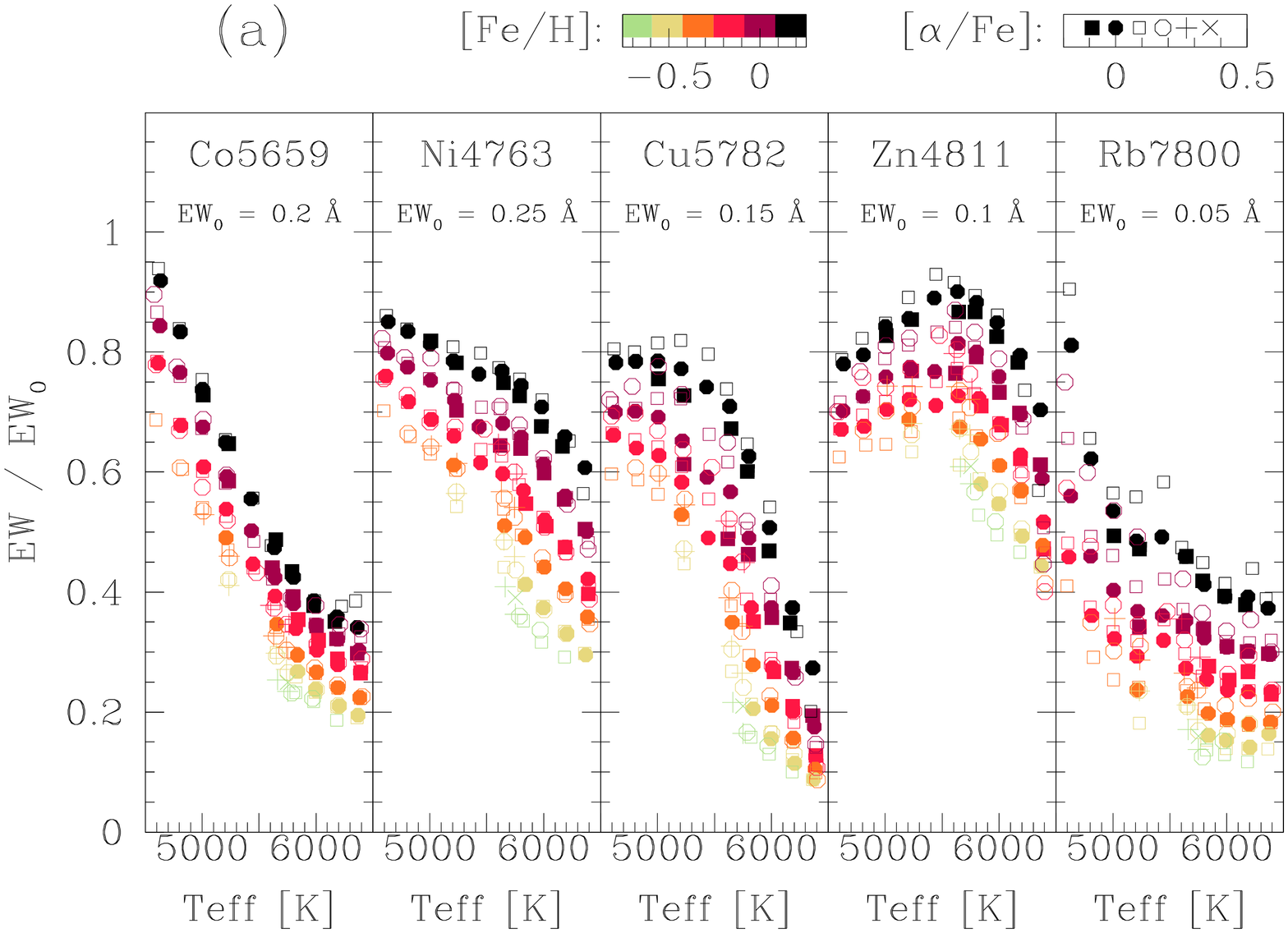} \\                
  \includegraphics[width=0.78\columnwidth,angle=270]{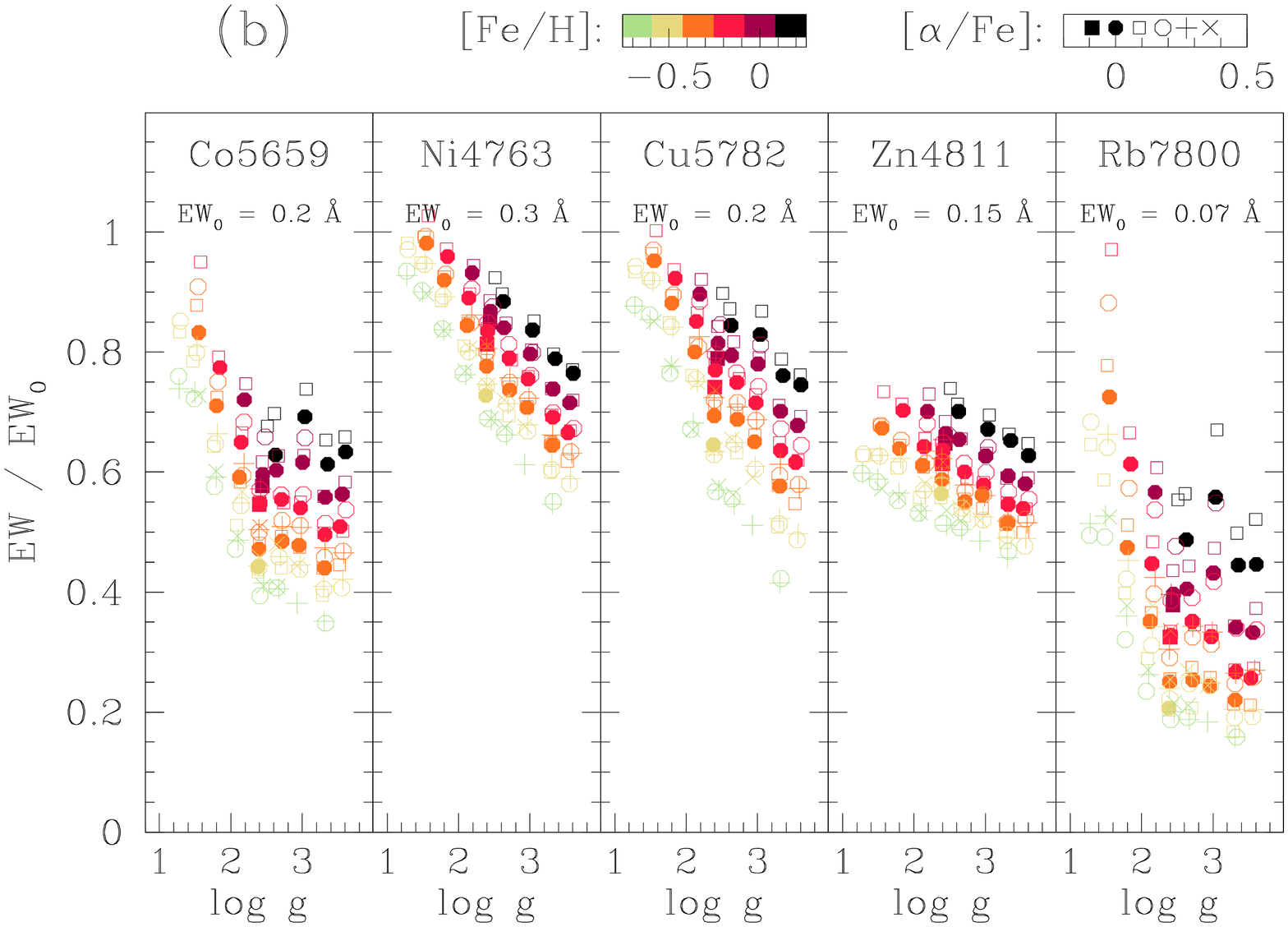}                 
  \caption{As Figure \ref{Figmed1}, but for lines of chemical elements with $27 \leq Z \leq 37$.}
  \label{Figmed4}
\end{figure}

\begin{figure}
  \includegraphics[width=0.78\columnwidth,angle=270]{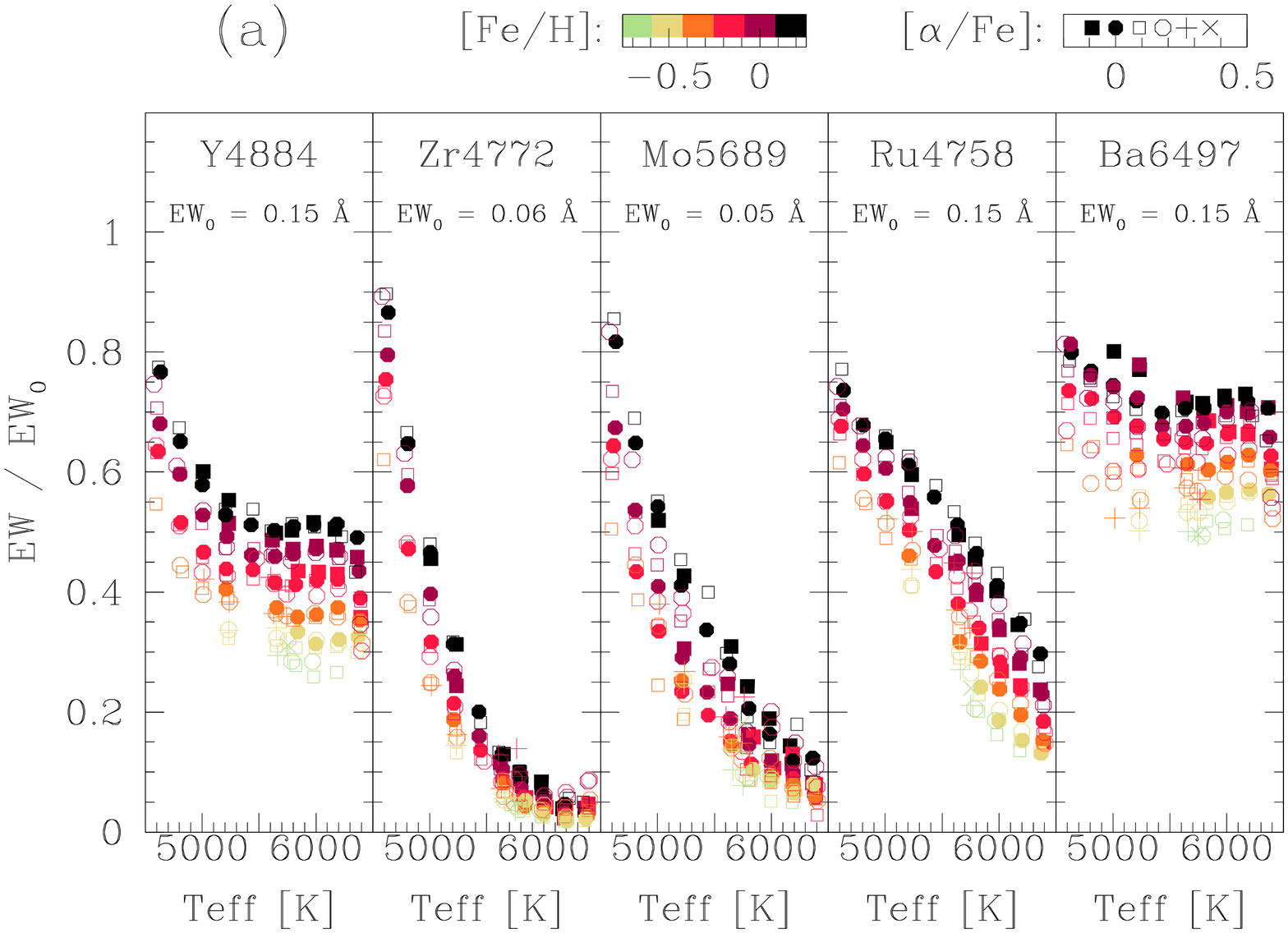} \\                
  \includegraphics[width=0.78\columnwidth,angle=270]{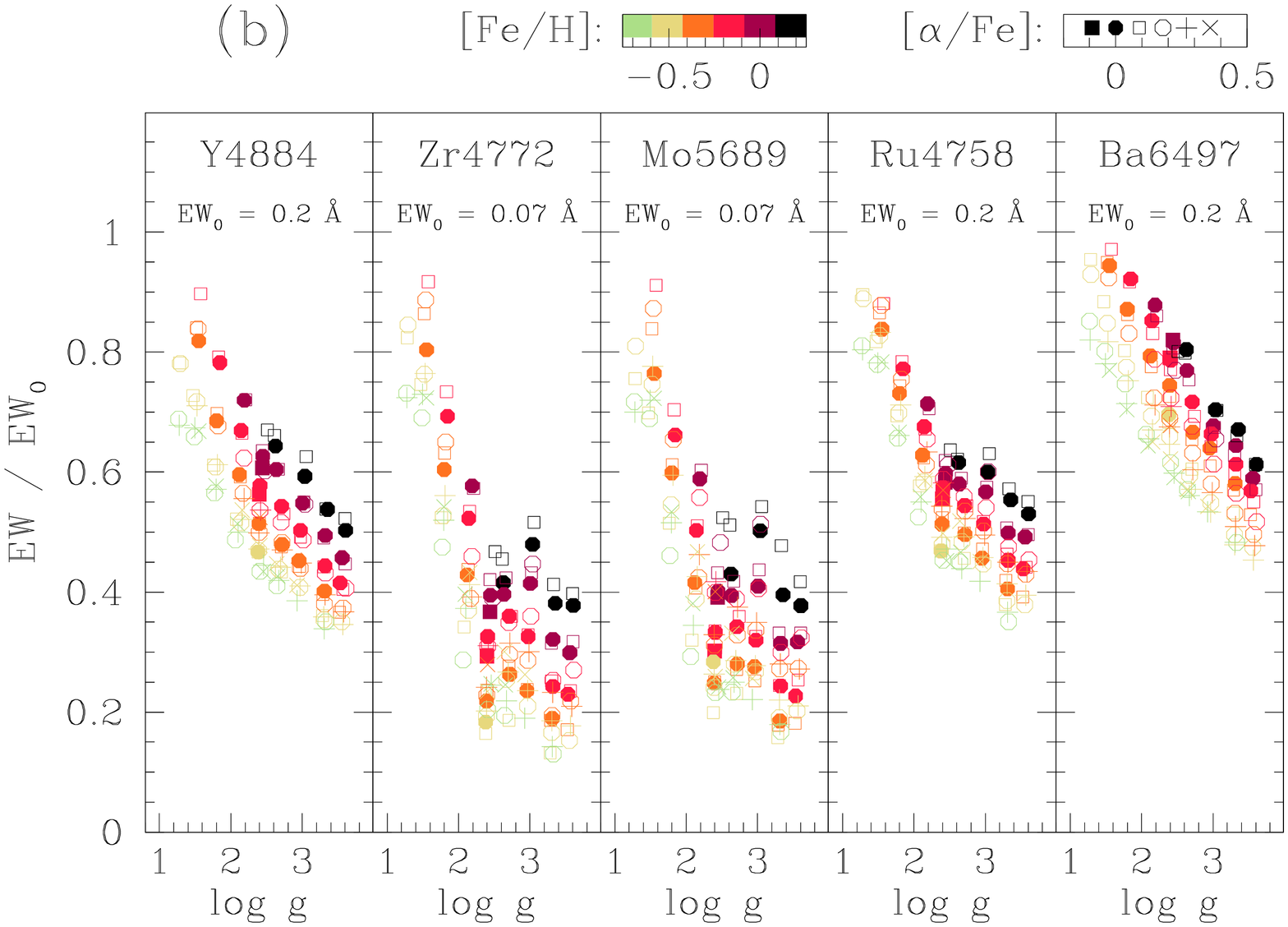}                 
  \caption{As Figure \ref{Figmed1}, but for lines of chemical elements with $39 \leq Z \leq 56$.}
  \label{Figmed5}
\end{figure}

\begin{figure}
  \includegraphics[width=0.78\columnwidth,angle=270]{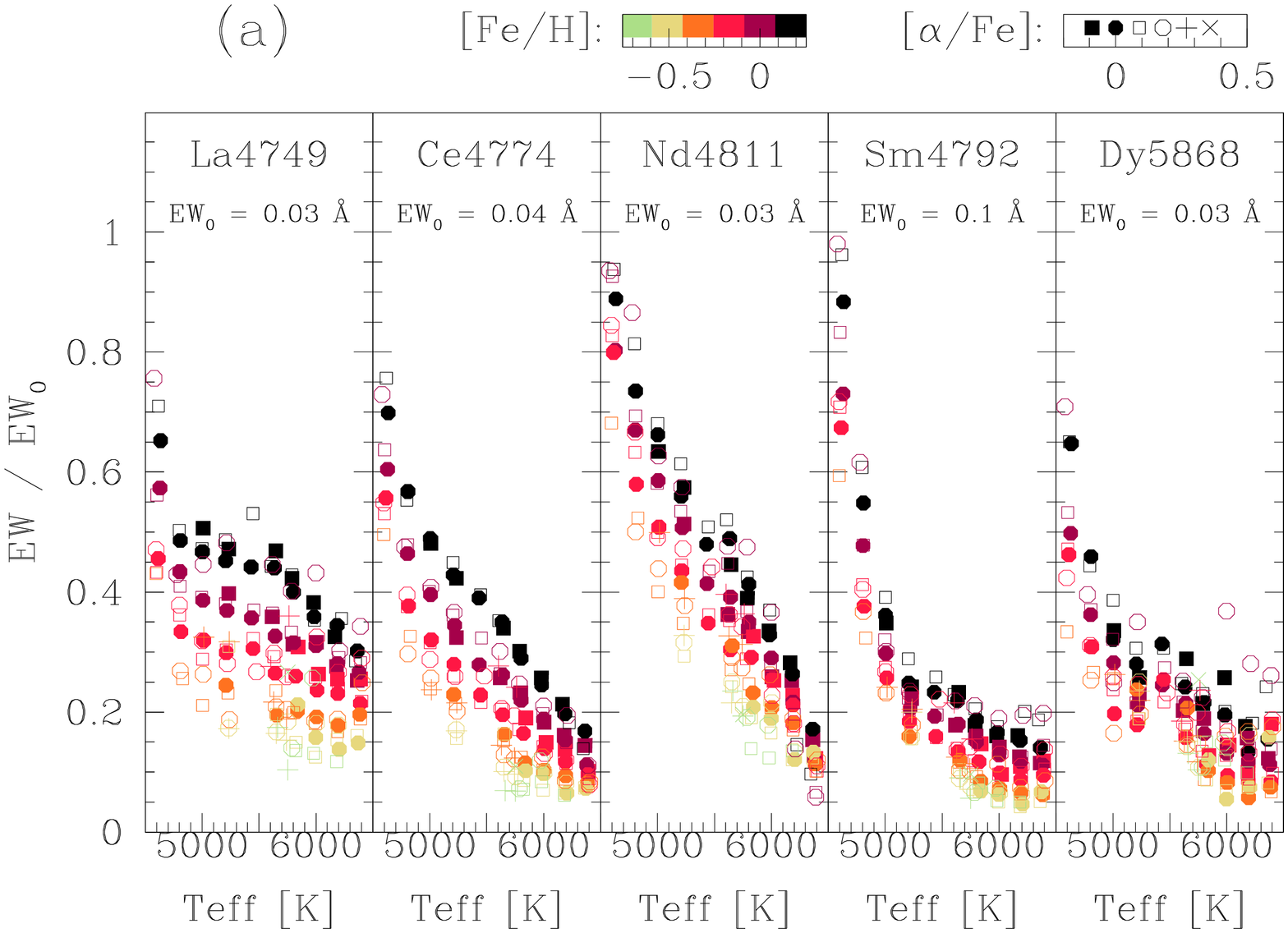} \\                
  \includegraphics[width=0.78\columnwidth,angle=270]{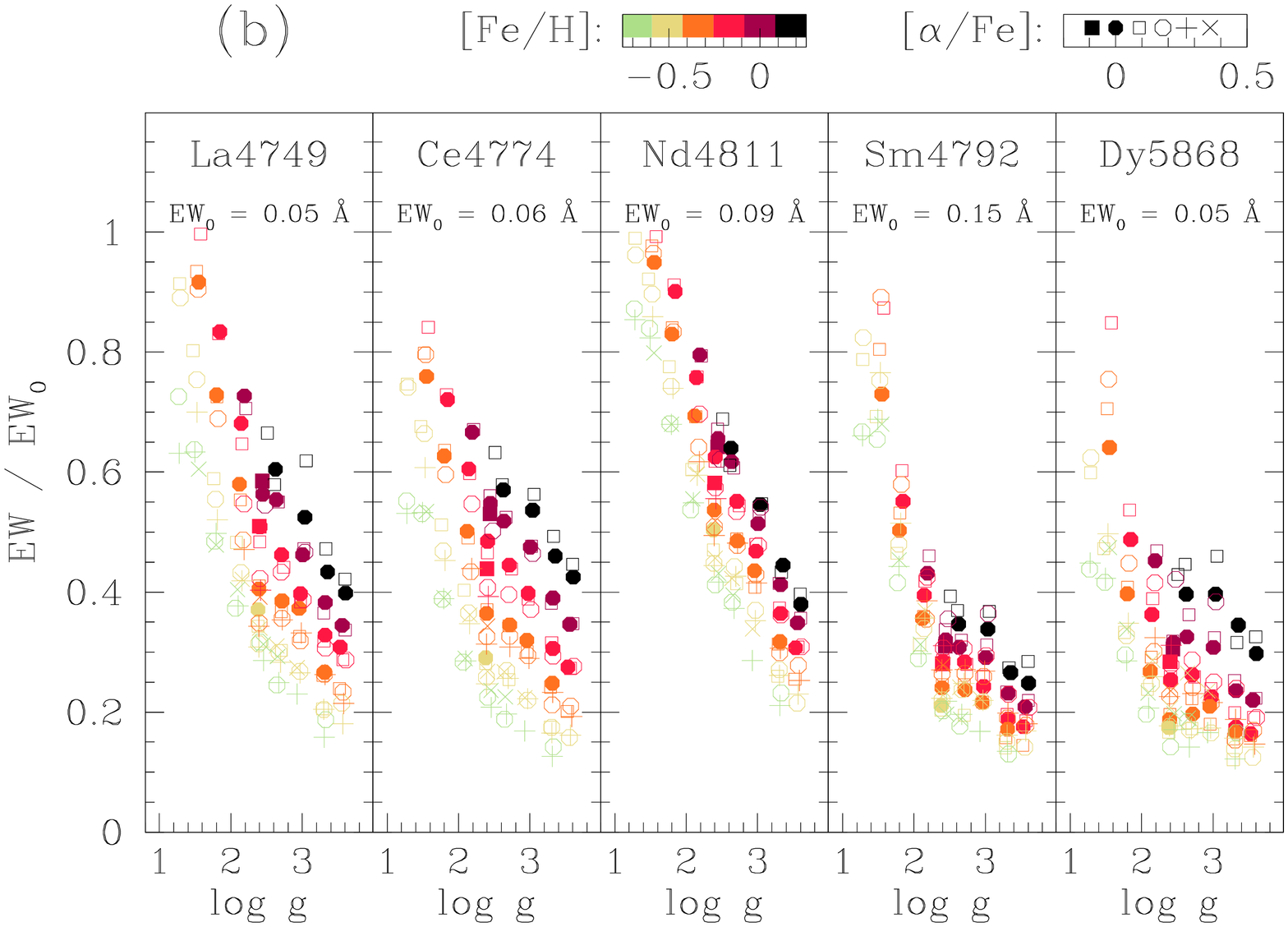}                 
  \caption{As Figure \ref{Figmed1}, but for lines of chemical elements with $57 \leq Z \leq 66$.}
  \label{Figmed6}
\end{figure}

A library of observed spectral templates can have different uses. An example is the measurement of equivalent widths (EWs) of individual spectral lines. This can be tricky in individual observed spectra, as noise does not allow an unambiguous placement of the continuum level. The situation with spectral medians is different, as they are nearly noise-free. Figures  \ref{Figmed1}--\ref{Figmed6} show EWs of the strongest unblended lines of 30 chemical elements measured by {\sl GALAH}, arranged by their atomic number $Z$. Reported EWs are calculated over a line mask range, with the continuum set to the maximum value within the segment mask range for each line (see Table A3 in \citetalias{buder20}). In each figure, the top panel shows EWs along the main sequence and the bottom one along the RGB, defined by grey bands in Figure \ref{FigKiel}. Iron abundances are colour-coded and $\alpha$ enhancements are presented with different symbols. As expected, iron group elements have all symbols (different $\alphafe$) with a given colour (a given $\feh$) overlapping, with a monotonic relation between EW and $\feh$. Since [X/H] correlates with [Fe/H] this explains correlations of lines that are not from the iron group. On the other hand EWs of $\alpha$ elements, especially along the RGB, have non-overlapping symbols of a given colour (different $\alphafe$ at the same $\feh$). The behaviour of some elements is entirely different, such as Li~I 6708, C~I 6588 and O~I 7772 (Fig.\ \ref{Figmed1}), Si~I 5684 (Fig.\ \ref{Figmed2}), Cu~I 5782 and Zn~I 4811 (Fig.\ \ref{Figmed4}), Ba~II 6497 (Fig.\ \ref{Figmed5}), and Sm~II 4792  (Fig.\ \ref{Figmed6}). Discussion of individual cases is beyond the scope of this paper. These graphs demonstrate the well known fact that the strength of spectral lines can serve as a sensitive stellar thermometer, and that medians of observed spectra present an interesting overview of spectral changes across the HR diagram. 

\section{RV measurement pipeline}
\label{SecRVs}

As a first step we compute RV shifts of each observed spectrum versus the relevant observed median spectrum. This is done as in \citetalias{zwitter18}, using an iterative process of computing a weighted average of 20 wavelength intervals along the four spectrograph arms. Next,  we need to compute the RV shift between the observed median spectrum and a suitable synthetic spectral library. We use the one of \citet{chiavassa18}, which includes three-dimensional convective motions within the stellar atmosphere. It has been computed using the radiative transfer code Optim3D \citep{chiavassa09} for the STAGGER grid of three-dimensional radiative hydrodynamical simulations of stellar convection \citep{magic13}. The convective motions give rise to convective blue-shifts which vary from line to line, depending on line strengths, element, ionisation stage, excitation potential etc.\ \citep[see e.g.][ for the Sun]{asplund00}. These are accounted for in the Chiavassa et al.\ 3D synthetic spectra which should improve the RV determinations.

Comparison of observed median spectra to synthetic ones can yield more than one RV measurement. In particular, one expects that RVs measured over different wavelength intervals of a given spectrum are consistent within errors. If they are not and if deviations in a given wavelength region are seen over a range of spectral types this indicates a problem with the wavelength calibration of this region. This can be due to a lack of suitable ThAr calibration lines with accurately known wavelengths or a result of PSF variation, which is typical for fast focal-ratio spectrographs, including HERMES \citep{kos18}. 

\begin{figure}
  \includegraphics[width=0.78\columnwidth,angle=270]{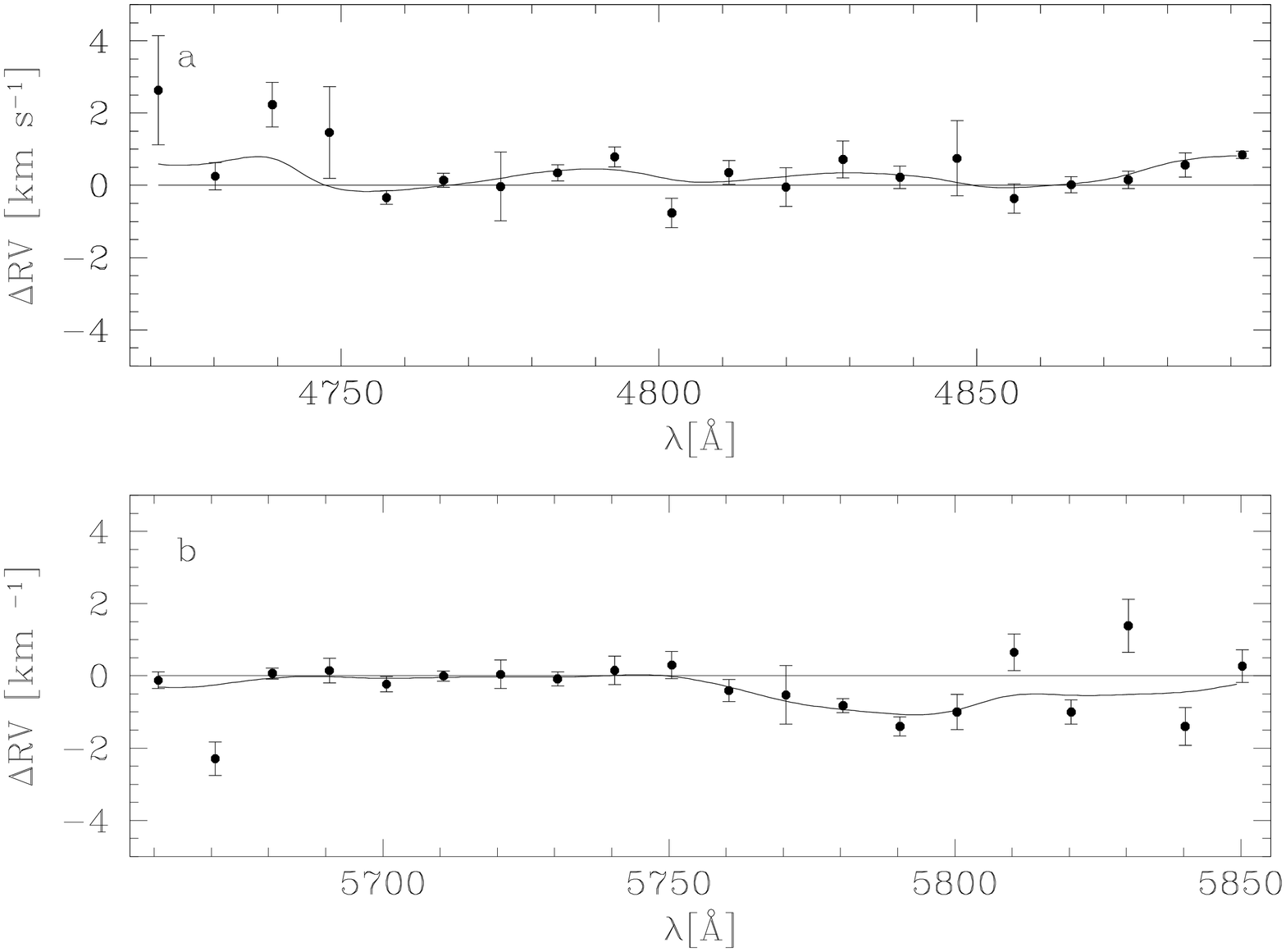} \\                
  \includegraphics[width=0.78\columnwidth,angle=270]{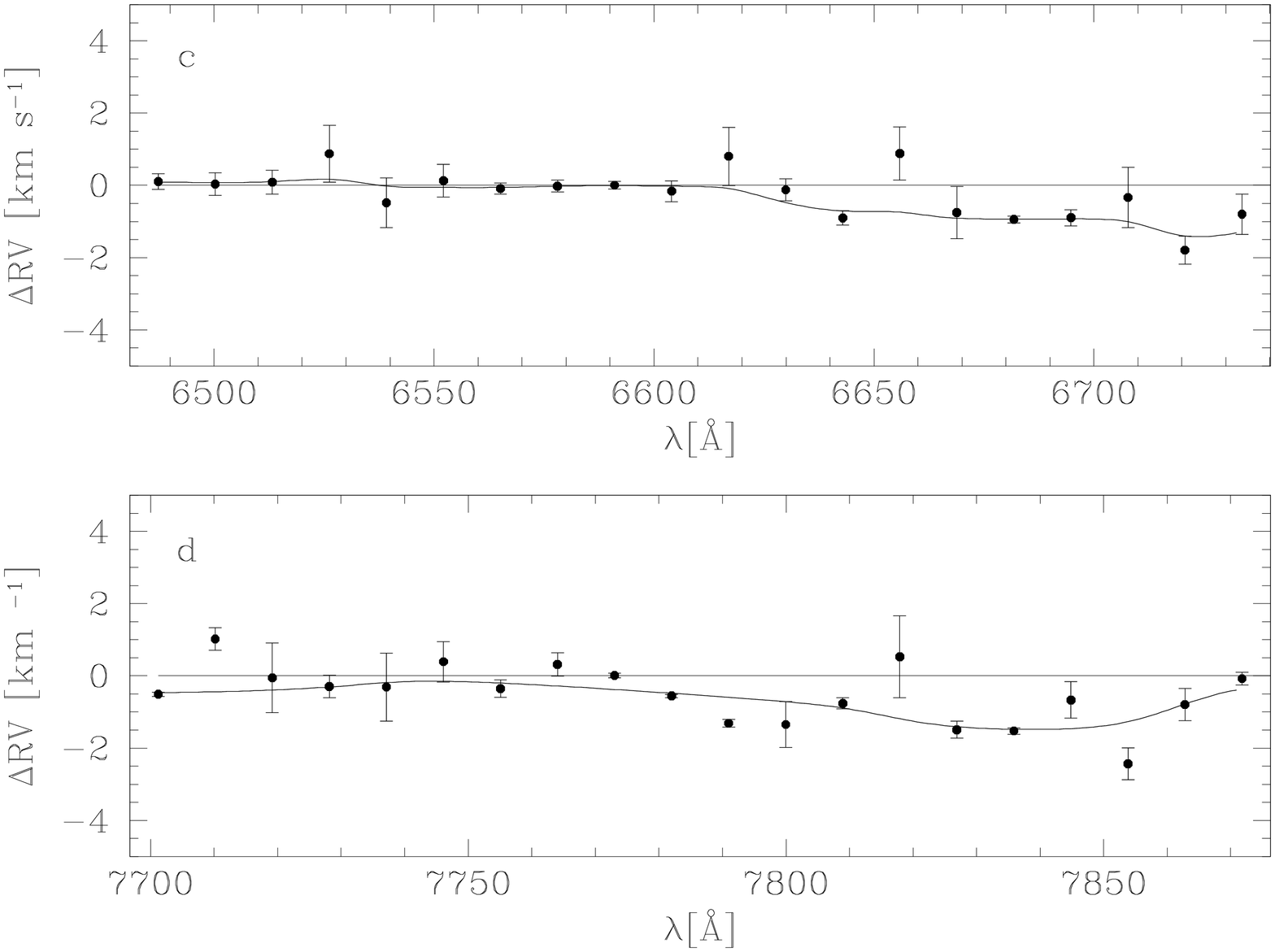}                 
  \caption{Correction to a wavelength solution as derived from a piecewise comparison of observed median spectra and synthetic templates. $\Delta$RV is the difference between a weighted average of median observed spectra and the overall RV, which is defined as a peak of a combined correlation function of all wavelength bins and over all arms of the spectrograph.}
  \label{FigRVshift}
\end{figure}

The RV calculation can be sensitive to a moderately different chemical composition of the observed median and synthetic spectra. So we first renormalise both spectra in each arm of the spectrograph using a 3-piece cubic spline with asymmetric rejection (low\_rej~$=3.0$, high\_rej~$=5.0$) and 10 iterations. To detect any systematic RV shifts we divide the spectral range of each arm into 20 fine intervals with a width of 10--14~\AA. This gives us 80 RV measurement points over 718 observed median templates. Figure \ref{FigRVshift} shows that some systematic shifts are present. We use 85 median templates which are based on the largest numbers of combined observed spectra and have $\teff \leq 6000$~K. We then compare their RV within a given wavelength bin to a peak of a combined correlation function of all wavelength bins and over all 4 arms of the spectrograph. In an ideal case we would expect a random scatter around zero. Figure \ref{FigRVshift} shows that some wavelength bins have large error-bars because they do not contain any strong spectral lines. Some points also show large offsets, which reflect a mismatch between synthetic and observed spectra or a presence of a strong not fully matched spectral line at the edge of the wavelength bin. All this is expected. But Figure \ref{FigRVshift} also shows that for example the red edges of the green, red and IR arms have residuals with consistently negative sign, indicating a systematic blueshift. 

The continuous curves in Figure \ref{FigRVshift} are derived as running weighted averages of individual points, penalising their wavelength distance by a Gaussian with $\sigma = 500$~\kms. This value was chosen to mimic a typical density of ThAr lines and an expected spatial variation of the PSF within the spectrograph. These curves are used as corrections to shift the RVs from the individual wavelength bins, which would help us deriving more consistent RVs of the observed median spectra. In particular, the typical RV uncertainties reported by Z18 were 0.09~\kms. The use of better values for the stellar parameters, the use of $\alphafe$ values for constructing the median observed spectra, a larger number of RV bins and a better implementation of the correlation routine now allows us to bring this down to 0.042~\kms. Finally, taking the just mentioned wavelength correction into account reduces the uncertainty to 0.027~\kms.  

A meaningful calculation of median spectra requires well populated bins in the ($\teff$, $\logg$, $\feh$, $\alphafe$) space. As discussed above there are 718 such bins, each with at least 100 un-flagged spectra. Together they contain 474,309 spectra, the other 106k spectra belong to less populous bins. For the less populous bins we use the median observed spectra from the closest well populated bin. That bin is determined as the one with the smallest Manhattan distance $\frac{\Delta \teff}{a} + \frac{\Delta \logg}{b} + \frac{\Delta \feh}{c} + \frac{\Delta \alphafe}{d}$, with $(a,b,c,d) = (2\,\mathrm{K}, 0.01\,\mathrm{dex}, 0.1\,\mathrm{dex}, 0.4\,\mathrm{dex})$. This choice of constants, which was found by trial and error to pick the most similar rescaled spectrum, reflects the fact that spectra change quickly with temperature, less with gravity, while chemistry mostly reflects only the depth of (certain) lines and thus has a minor influence on derived RVs.

\begin{figure}
  \includegraphics[width=0.78\columnwidth,angle=270]{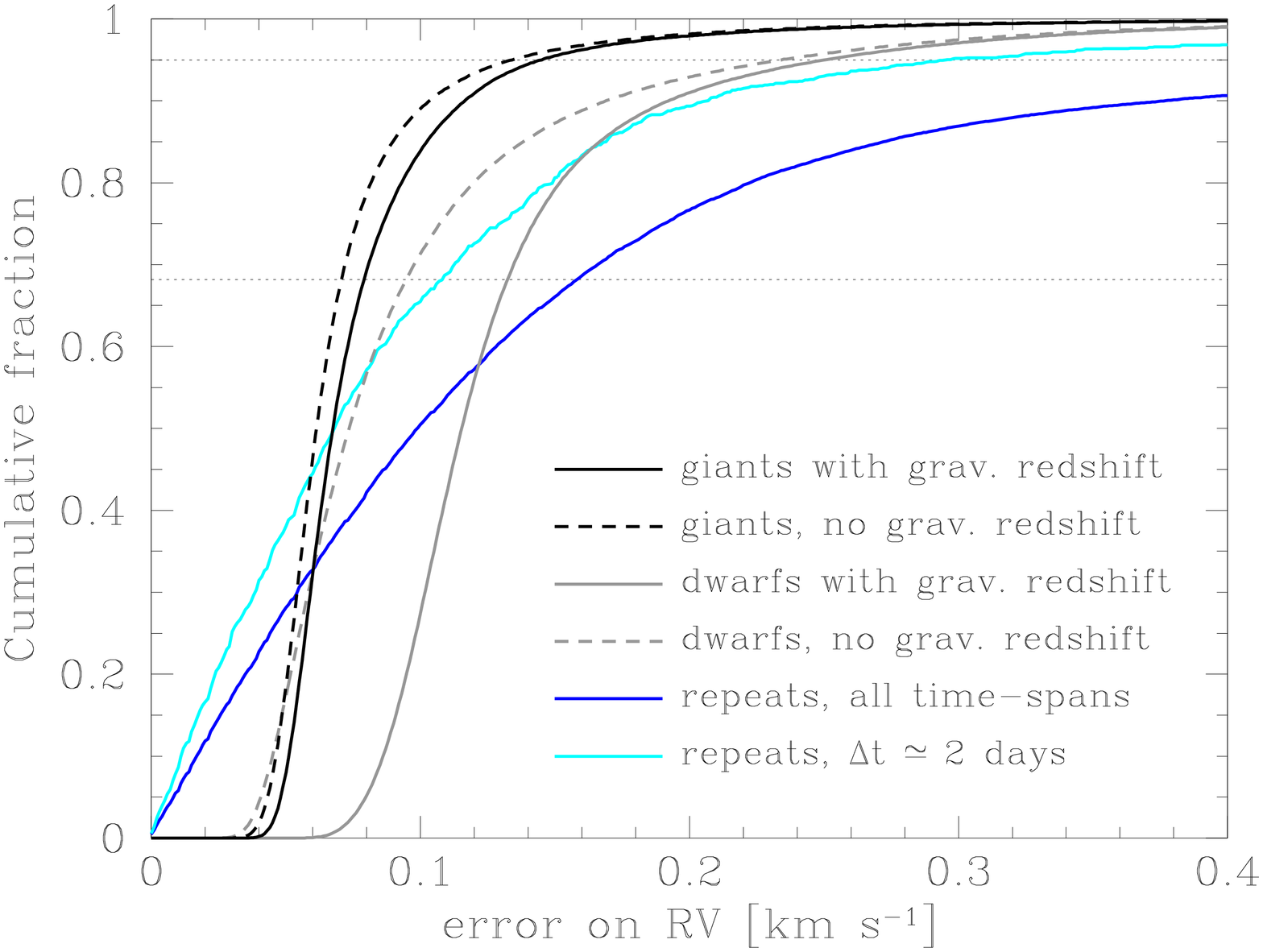}                 
  \caption{Cumulative histogram of formal radial velocity uncertainties for giants (black) and dwarfs (grey) with (full line) an without (dashed line) gravitational redshift correction. Coloured curves are cumulative distributions of standard deviation of actual repeated RV measurements of the same objects re-observed 2 days apart (cyan) or at
any time-span (blue). Horizontal dotted lines mark the 68.2 per cent and 95 per cent levels.}
\label{Figrverrors}
\end{figure}

To complete the RV calculation we need to consider two final steps. The first one is a barycentric correction. Here, it is done with the routine {\sl bcvcorr}, which is part of the {\sl IRAF} {\texttt{RVSAO} package and is more accurate than {\sl rvcorr} used in \citetalias{zwitter18}. Next, we note that light suffers from gravitational redshift as it travels from the stellar surface to the observer. This effect is substantial, it reaches 0.636~\kms\ for a solar type star and is proportional to the ratio of the stellar mass and radius. So stars of different types would show inconsistent RVs if this effect was neglected. 

The implementation of gravitation redshift is the same as described in \citetalias{zwitter18}, but benefits from better values of the stellar parameters. Still, one should note that the radius of the star is difficult to determine accurately. So the final velocities, which take gravitational redshift into account, are internally consistent but have substantially larger uncertainties than without taking gravitational redshift into account. So one should use values corrected for gravitational redshift if different types of stars are to be compared, such as within a stellar cluster or in studies of Galactic dynamics. But if the goal is to study RV variability of a certain star, the values without gravitational redshift correction are preferred because of the more realistic uncertainties. Also, most stars can be regarded as static in our sample. However, we measure the stellar parameters for each spectrum separately even when multiple spectra per star are available, which may induce some variation in calculation of gravitational redshifts. Finally, most of the published catalogues, including RVs from {\sl Gaia}, do not include gravitational redshift corrections at present. So in general, one needs to use our values without gravitational redshift correction to compare them to the ones from the literature. As explained below we therefore publish RVs with and without gravitational redshift correction. 

Figure \ref{Figrverrors} plots cumulative distributions of uncertainties for RVs with and without gravitational redshift correction, separately for dwarfs and giants (the dwarf-giant separation line is defined in eq.\ 1 in \citetalias{zwitter18}). The uncertainties were derived using a strict error propagation, as explained in \citetalias{zwitter18}. The values are now  $\sim 25$\%\ smaller, mostly due to more accurate stellar parameters, a better wavelength calibration, and other computational improvements, as explained above. The gravitational redshift correction increases the uncertainties, because there are uncertainties in the determination of the stellar parameters that need to be factored in. Values for giants are better than for dwarfs, which reflects the abundance of spectral lines and smaller importance of the gravitational redshift correction for giants compared to dwarfs. We see that the majority of uncertainties, especially when considering measurements without the gravitational redshift correction, are within 0.1~\kms.

Table \ref{tableRVs} reports results of the RV measurement pipeline. The whole table is available electronically. 

\begin{table*}
\caption{All RVs presented in this paper forming a Value added catalogue of RVs of {\sl GALAH$+$} DR3.  The velocity columns list the RV and its uncertainty for measurements including the gravitational redshift (RV) and for those without this correction (RV\_nogr). MJD is the local modified Julian date, and JD the heliocentric Julian date. A complete list is published electronically. }
\begin{tabular}{cccrrcc} 
sobject\_id&2MASS\_id&Gaia\_DR2\_id&\multicolumn{1}{c}{RV}&\multicolumn{1}{c}{RV\_nogr}&MJD&JD\\ 
 & & &\multicolumn{1}{c}{\kms}&\multicolumn{1}{c}{\kms}&(local)&(heliocentric)\\ \hline
131116000501002&03325271-6840304&4667368899326729856&36.204$\pm$0.183&36.890$\pm$0.138&56612.5155509&2456613.01580\\
131116000501004&03422255-6841522&4667324643983679744&95.878$\pm$0.151&95.914$\pm$0.150&56612.5155509&2456613.01579\\
131116000501005&03373408-6841062&4667335913977929728&7.130$\pm$0.126&7.565$\pm$0.096&56612.5155509&2456613.01579\\
131116000501006&03430488-6843208&4667323681911007232&24.816$\pm$0.163&25.349$\pm$0.159&56612.5155509&2456613.01579\\
131116000501007&03425716-6844462&4667323544472053888&-38.360$\pm$0.136&-37.917$\pm$0.099&56612.5155509&2456613.01579\\
... & ... & ... & ... & ... & ... & ... \\\hline
\end{tabular}
\label{tableRVs}
\end{table*}

\section{Objects with variable and with constant RVs} 
\label{Secrepeats} 

{\sl GALAH$+$} is mostly a single visit survey. But 25,358 stars have more than one spectrum satisfying flag\_sp~$=0$ and have RVs measured by our procedure. From these 21,379 have a pair of observations, 2784 have 3 visits, 676 have 4 visits, 203 have 5 visits, 139 have 6 visits, 125 have 7 visits, 51 have 8 visits, and one star has been observed 9-times. Such a statistics of repeated observations is not enough to study properties of stars with variable RVs or to determine if a star has a stable RV over a long term. Still, with a suitably stringent selection criterion, we can identify candidates with intrinsically variable RVs and candidates with constant RVs. In both cases we use the RVs without gravitational redshift correction, as this correction would inflate the RV error-bar due to uncertainties in the values of the stellar parameters. 

\begin{table*}
\caption{Stars with variable RVs that exceed a 100~\kms\ amplitude. $N$ is number of observations. The last 4 columns give details of the pair of measurements with the largest RV difference. A complete list of 4483 stars with variable RV at a $4\,\sigma$ level is available electronically. }
\begin{tabular}{cccrrcc} 
2MASS\_id &Gaia\_DR2\_id&N&$|RV_2-RV_1|$&  $t_2-t_1$ &       sobject\_id$_1$ &       sobject\_id$_2$  \\ 
                   &                        &   & \kms               & days & & \\ \hline
09541851-6939098&5243109471519822720&2& 196.756&  737.98624&151225004301112&180101005001112\\
10015471-4131014&5418823008865449472&2& 147.292&  237.37062&170507006201267&171230006301071\\
05203528+0120357&3234152606303000576&3& 123.536&    3.02859&190209002401182&190212002001182\\
04071697-6301357&4676308341177919744&7& 123.487&  246.31468&170107001801301&170910005601301\\
07050942-6724015&5280932808950813824&2& 122.571&  710.89050&170110002101048&181222003601047\\
04111667-6739489&4668410652234512256&4& 112.460&    0.91227&171207002701149&171208001601149\\
06152332-6159006&5481076085919696640&2& 105.065&    2.84710&170203001901305&170206002901305\\
06475902-3407140&5582458995101542656&2& 104.242& 1393.17875&140312001701146&180103003101146\\
... & ... & & ... & ... & ... & ... \\\hline
\end{tabular}
\label{tablevariableRV}
\end{table*}

\begin{table*}
\caption{Stars with constant RVs with at least $N\geq 3$ observations that span $\Delta t \geq 1$~year in time and $\Delta RV < 0.2$~\kms\ in their individual RV\_nogr measurements. The last two columns list the weighted average RV and its uncertainty for measurements including the gravitational redshift (RV) and for those without this correction (RV\_nogr).  A complete list of 225 stars with constant RVs is available electronically. }
\begin{tabular}{ccclrrclrcl} 
2MASS\_id &Gaia\_DR2\_id&N&$\Delta RV$&  $\Delta t$ &\multicolumn{3}{c}{RV}&\multicolumn{3}{c}{RV\_nogr}\\  
                   &                        &  & \kms           & days          &\multicolumn{3}{c}{\kms}&\multicolumn{3}{c}{\kms}\\ \hline 
04062738-6252547&4676358403316733696&8&   0.170& 1476.82127&  37.108&$\pm$&   0.053&  37.458&$\pm$&   0.049\\
04111980-7051077&4654280897025535360&4&   0.044&  506.65084&  10.582&$\pm$&   0.055&  10.997&$\pm$&   0.047\\
12044223-3949215&3459350489096016640&4&   0.047&  441.75844&  24.948&$\pm$&   0.024&  25.002&$\pm$&   0.024\\
12061470-3944312&3459344132544717824&4&   0.110&  443.79997&  61.912&$\pm$&   0.054&  61.969&$\pm$&   0.054\\
12051308-4010565&3458953393599646336&4&   0.146&  443.79996&  14.260&$\pm$&   0.059&  14.299&$\pm$&   0.057\\
12064427-4009095&6149476076392742144&4&   0.159&  441.81152&  67.091&$\pm$&   0.069&  67.609&$\pm$&   0.063\\
08233294-1919135&5707436496101210624&4&   0.160&  387.00457&  33.492&$\pm$&   0.090&  34.120&$\pm$&   0.064\\
04111504-7139337&4653848406703013760&4&   0.163&  831.78819& --5.834&$\pm$&   0.079& --5.272&$\pm$&   0.068\\
04094116-6317034&4676286621527513344&4&   0.166&  384.94644& --6.122&$\pm$&   0.070& --5.500&$\pm$&   0.063\\
12035022-3947097&3459023762343865984&4&   0.184&  441.75841&  59.808&$\pm$&   0.062&  59.912&$\pm$&   0.063\\
04120308-6114296&4676931386313930496&4&   0.198& 1476.85275&  75.390&$\pm$&   0.098&  75.493&$\pm$&   0.093\\
... & ... & & ... & ... &\multicolumn{3}{c}{...}&\multicolumn{3}{c}{...}\\\hline
\end{tabular}
\label{tableconstantRV}
\end{table*}

To quantify the significance of RV differences between two measurements with assumed Gaussian distributions  ($RV_1$, $\sigma_1$) and ($RV_2$, $\sigma_2$) one can write the probability $P$ that a random pick from the second distribution is larger than the one from the first one:
\begin{equation}
P= 
\frac{1}{2\pi \sigma_1 \sigma_2} 
\int_{-\infty}^\infty \int_{-\infty}^{y} \exp{\frac{-(x-RV_1)^2}{2 \sigma_1^2}} \exp{\frac{-(y-RV_2)^2}{2 \sigma_2^2}} {\mathrm d}x {\mathrm d}y
\end{equation}
which can be simplified to \citep{matijevic11}
\begin{equation}
P= \frac{1}{2}\left[ 1+\mathrm{erf}\left(\frac{|RV_1-RV_2|}{\sqrt{2(\sigma_1^2+\sigma_2^2}}\right) \right] 
\end{equation}
where $\mathrm{erf}$ is the standard error function. For two measurements with nearly matching RVs this probability is close to $\frac{1}{2}$, but for an object with a significant RV variation the value will converge to 1. If we use a $4\,\sigma$ type of criterion for the detection of RV variability, hence $P > 0.9999366575$, we find 4483 stars with variable RV. From these 2592 are main sequence objects (defined as in \citetalias{zwitter18}). Intrinsically variable stars are listed in Table~\ref{tablevariableRV} to be published in full in electronic form at the CDS.  

In most cases the number of RV observations is too small to derive a solution of the RV curve. But, as said, there are 177 stars with 7 or more observations. For example, the star 2MASS $04071697-6301357 \equiv \mathrm{Gaia\,DR2}\,4676308341177919744$ has 7 observations spanning a RV range of 123.487~\kms\ (Table~\ref{tablevariableRV}). A circular solution $RV = K \sin{2 \pi \frac{t-t_0}{P}}+\gamma$, with $K = 65.36(1)$~\kms, $P=3.3488673(28)$~days, $\gamma=9.233(1)$~\kms, $BJD(t_0)=2457571.1256(2)$, fits the observations with  $\overline{O-C}=0.339$~\kms. Its mass function equals $0.0969\mathrm{M_\odot}/\sin^3 i$ and is consistent with a low mass secondary component. The star has not been studied so far, except by the {\sl 2MASS} \citep{cutri03} and {\sl Gaia} \citep{brown20} surveys. {\sl Gaia} reports a large uncertainty of mean RV, which is compatible with our results. 

Another important use of repeated observations is to establish which stars have a constant RV and can therefore serve as RV standards. {\sl GALAH$+$} generally observes fainter stars than listed in published catalogues of RV standards. But the number of {\sl GALAH} observations is generally too small and their time-span too short to firmly establish them as RV standards. Still, we can build a list of candidates, which can be used for validation of other surveys, though with a caveat that some of the targets may eventually turn out to have a variable RV. Table \ref{tableconstantRV} lists 225 objects with at least 3 observations which span more than a year in time and less than 0.2~\kms\ in their individual RV measurements (without gravitational correction).  

Repeated observation can be used to verify the precision of the derived RVs. The blue curve in Figure~\ref{Figrverrors} is a cumulative histogram of all pairs of measurements of the same objects at all time-spans. Note that many of these objects are intrinsically variable, as discussed above. So the cyan curve shows results for repeated observations obtained 2 days apart, which should filter out any long-term variability. It shows that 68.2\%\ of the pairs of measurements are within $\pm 0.109$~\kms. Use of other short time-spans yields similar results. The exception is observation within the same night or in consecutive nights, which have typical uncertainties of  $\pm 0.127$~\kms. In such cases the second observation was often obtained because of an unsatisfactory quality of the first one.

\section{Motions within M67}
\label{SecM67}

A combination of {\sl Gaia} astrometry and our RVs allows us to study the three-dimensional position and velocity vectors of stars {\sl within} stellar streams or clusters. Here we use the open cluster M~67 as an example. The same approach can be expanded to other clusters and associations. 

We selected M~67, as {\sl GALAH$+$} observed 244 of its members listed in \citet[][hereafter C19]{carrera19}. Its age is 3.64~Gyr \citep{bossini19}, with recent estimates ranging from 3.46~Gyr \citep{stello16} to 4.2~Gyr \citep{barnes16}, so its stars do not show signs of activity typical for young objects, which may otherwise complicate RV determinations. Its old age means it can be assumed to be dynamically relaxed from its birth motions. Its distance is typical for stars observed by {\sl GALAH} (the median parallaxes of {\sl GALAH} stars is 1.18~mas), so the results can be similar for other streams or associations. 


{\sl GALAH$+$} observed only  a quarter of the cluster's members, so general properties of M~67 need to be adopted from the literature. We assume the cluster's centre is at $\alpha_c = 132.84595^\mathrm{o}$, $\delta_c =11.813988^\mathrm{o}$  (epoch 2016.0), its proper motion is $\mu_{\alpha c} = -10.986$~mas/yr, $\mu_{\delta c}  = -2.964$~mas/yr, and the distance to the cluster centre is $d_c=860$~pc  \citep{cantat18}. 

Next, we define the coordinate system. A star at a distance $d$, with equatorial coordinates $(\alpha,\delta)$, proper motions $(\mu_\alpha,\mu_\delta)$ and radial velocity $\mathrm{RV}$ can have its position written in Cartesian coordinates as $d (\cos \delta \cos \alpha, \cos \delta \sin \alpha, \sin \delta)$. For convenience we translate and rotate this system, so that it is centred on M67, has its $z$ axis pointing away from Earth, while the $x$ an $y$ axes are tangential to the celestial sphere and pointing to the east and north, respectively. The position of the star $\vec{r} = (x, y, z)$ now becomes 
\begin{eqnarray}
x&=& d \cos \delta \sin(\alpha - \alpha_c) \\
y&=& d [\sin \delta \cos \delta_c -\cos \delta  \sin \delta_c \cos (\alpha - \alpha_c)] \notag\\
z&=& d [\sin \delta \sin \delta_c + \cos \delta \cos \delta_c \cos (\alpha - \alpha_c)]  - d_c \notag
\end{eqnarray}
and its velocity vector $\vec{v} = (\dot{x},\dot{y},\dot{z})$  with respect to the centre of the cluster is  
\begin{eqnarray}
\dot{x}&=&\mathrm{RV} \cos\delta \sin(\alpha - \alpha_c) - d \mu_\delta \sin\delta \sin(\alpha - \alpha_c) \\
& &+d \cos(\alpha - \alpha_c) (\mu_\alpha - \mu_{\alpha c} \cos\delta / \cos\delta_c )  \notag\\
\dot{y}&=&\mathrm{RV} [\sin \delta \cos \delta_c -\cos \delta \sin\delta_c \cos(\alpha - \alpha_c)] \notag\\ 
& &+ d \sin\delta \sin\delta_c [\mu_\delta \cos(\alpha - \alpha_c) - \mu_{\delta c} ] \notag \\
& &+ d \cos\delta \cos\delta_c [\mu_\delta  - \mu_{\delta c}\cos(\alpha - \alpha_c) ] \notag \\
& &+d \sin\delta_c \sin(\alpha-\alpha_c)[\mu_\alpha - \mu_{\alpha c} \cos\delta / \cos\delta_c] \notag\\
\dot{z}&=&\mathrm{RV} [\cos\delta \cos\delta_c \cos(\alpha-\alpha_c) + \sin\delta \sin\delta_c ] - \mathrm{RV}_c \notag\\
& & + d \sin\delta \cos\delta_c [-\mu_\delta \cos(\alpha-\alpha_c) + \mu_{\delta c}] \notag\\
& & + d \cos\delta \sin\delta_c [\mu_\delta - \mu_{\delta c} \cos(\alpha-\alpha_c)] \notag\\
& & + d \cos\delta_c \sin(\alpha-\alpha_c) [-\mu_\alpha + \mu_{\alpha c} \cos\delta / \cos\delta_c ] \notag
\end{eqnarray}
where $\mathrm{RV}_c$ is the RV of the cluster centre.  These cartesian coordinates are co-moving with the cluster centre ($\vec{r}_c=\vec{v}_c=0$), so they correct for perspective effects \citep{vandeven06}. 

For further study we select 193 stars that  are not in binary systems \citepalias[according to the classification by][]{carrera19}, have $\teff < 6150$~K (thus avoiding blue stragglers), have their parallaxes and proper motions published in \citet{brown20}  and which have all their RV measurements within 3~\kms\  of the $\mathrm{RV}_c$ (faster stars are probably unbound, unrelated to the cluster, or have intrinsically variable RVs). The value of $\mathrm{RV}_c$ was determined iteratively from our measurements so that the median of $\dot{z}$ is zero,  yielding $\mathrm{RV}_c =  +33.927 \pm 0.054$~\kms.  We always use RV values with a gravitational redshift correction, because only these yield consistent values for different stellar types. This result is similar to \citet{geller15} who derive the cluster's RV as $+33.64$~\kms\ from a much larger sample of RV measurements but neglecting convective and gravitational shifts. \citetalias{carrera19} classifies 9 of our stars as members of the red clump (RC), 25 as red giant branch (RGB) stars, 18 as subgiants (SGB), 65 as main-sequence turn-off (MSTO), and 76 as main-sequence (MS) stars. \citetalias{buder20} uses isochrone fitting to estimate their masses $m$. The  values span the range between 0.77 and 1.71~M$_\odot$, in agreement with a very flat mass function of M~67 \citep{hurley05}.

The position of the stars on the sky plane is known very well, but their distance has a larger uncertainty. The median uncertainty in the parallax of the cluster members, as listed by \citetalias{brown20}, is $19.3 \mu$as, which at the distance of M~67 translates to 14.3~pc. The astrometric precision of {\sl Gaia} is truly fantastic and unprecedented, but typical distance uncertainties are still comparable to the size of the cluster ($\sim 15$~pc), as determined by \citet{babusiaux18}. So we cannot use a simple inversion of the parallax to determine the distance of a star. Note that use of a general Galactic prior \citep{bailerjones18} is not appropriate due to a different space distribution of cluster members. We adopt a Nuker surface density profile \citep{vandermarel10}, with coefficients as derived by \citetalias{carrera19}, and invert it into a spherically symmetric radial probability distribution of cluster stars. This probability distribution is then sampled along the line-of-sight column of each of our stars. Such a procedure can determine a distribution of possible distances for each star which is then sampled with $10,000$ realisations, discarding outliers with a distance difference to the cluster centre larger than 17~pc \citep[an approximate limiting radius of the cluster,][]{gao18}. Similarly, we use reported uncertainties to sample RVs and proper motion values (taking their correlations into account) and determine a distribution of the three dimensional space coordinates and velocity vectors for each star. Individual realisations of the velocity vector for a given star have a small average spread ($\sigma_{\dot{x}} = 0.08$~\kms, $\sigma_{\dot{y}} = 0.06$~\kms, $\sigma_{\dot{z}} = 0.12$~\kms), the same is true for the $x$ and $y$ coordinates ($\sigma_x = 0.008$~pc, $\sigma_y = 0.008$~pc), while the spread in $z$ ($\sigma_z = 3.47$~pc) is much larger, though smaller than what would be obtained from a simple parallax inversion.

\begin{figure}
  \includegraphics[width=0.78\columnwidth,angle=270]{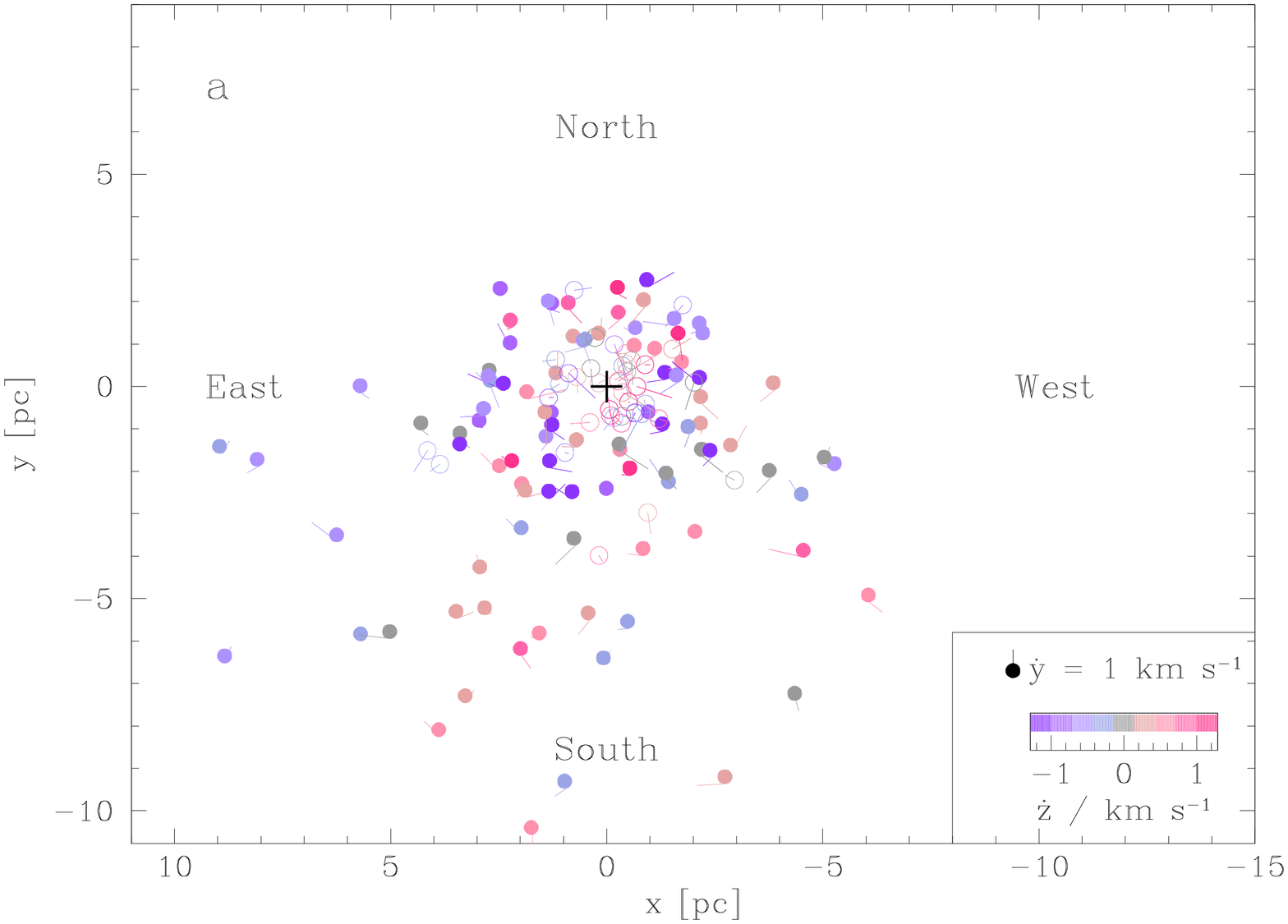}                 
  \includegraphics[width=0.78\columnwidth,angle=270]{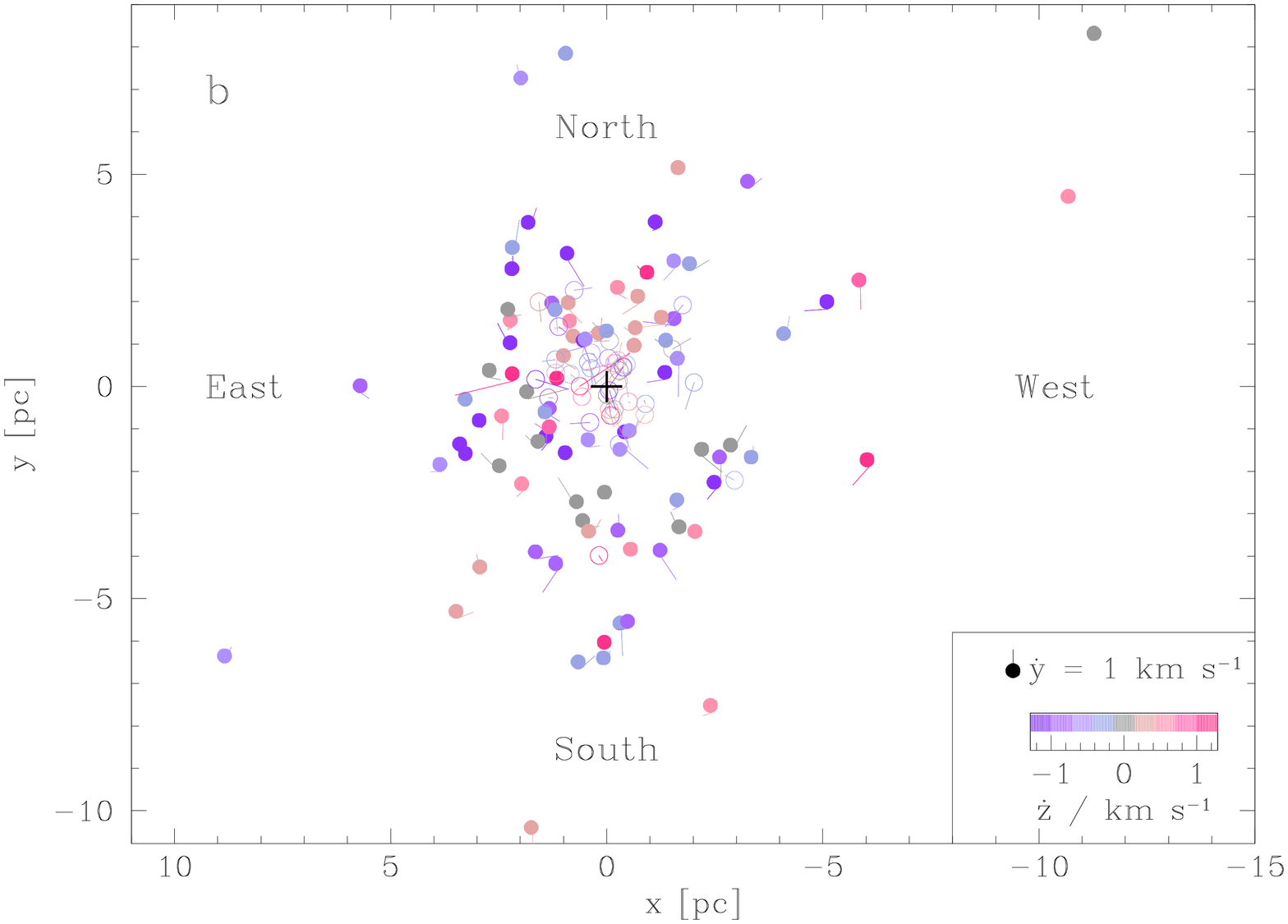}                 
  \includegraphics[width=0.78\columnwidth,angle=270]{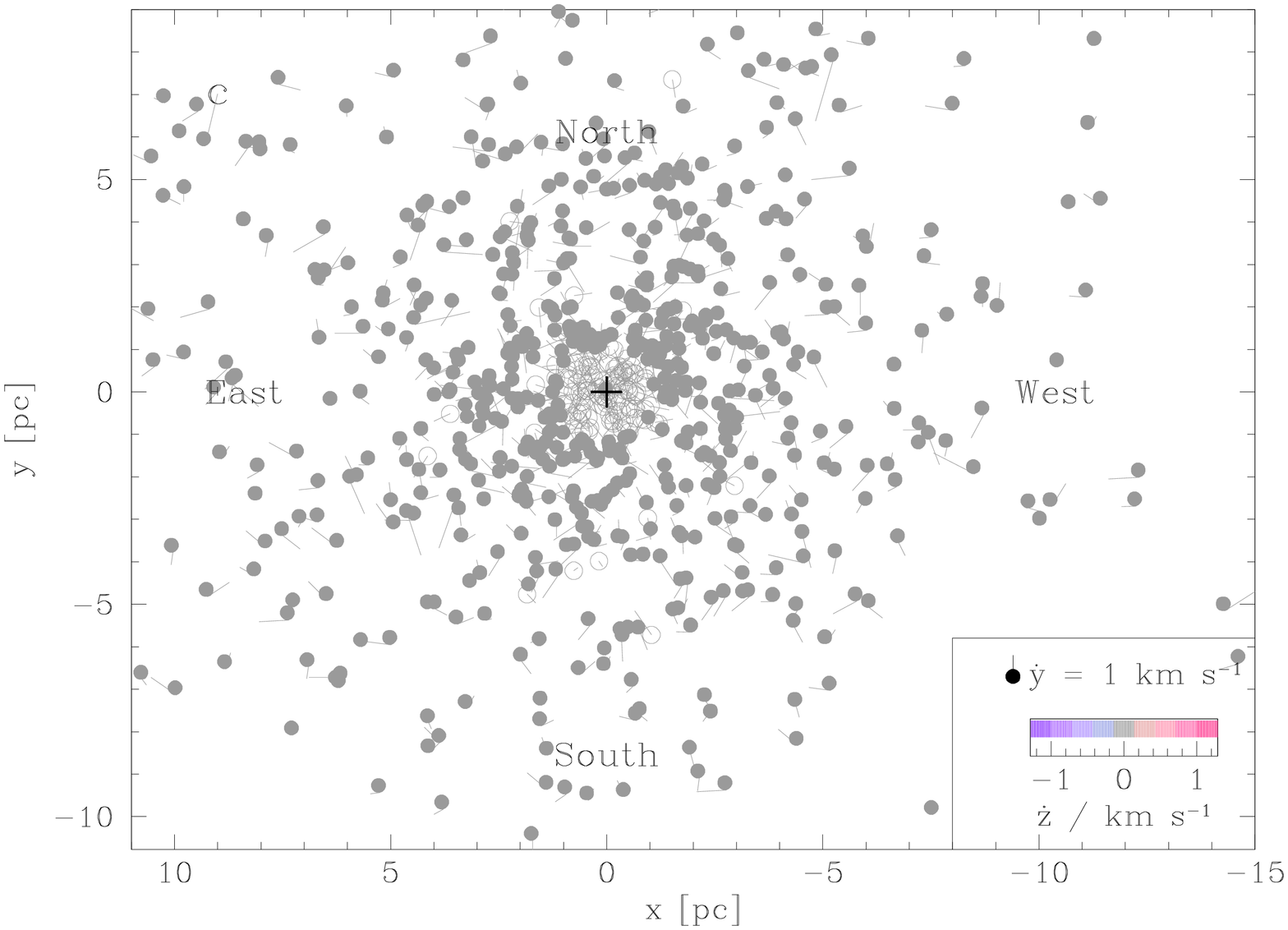}                 
  \caption{Positions and velocity vectors of single stars in M~67  that were observed  by {\sl GALAH$+$} (a), by {\sl APOGEE} DR16 (b), and by Gaia  eDR3 (c). In the latter case all RVs were assumed to be equal to $\mathrm{RV}_c$.  The position of each star is marked with a dot with a colour indicating its $\dot{z}$ velocity, and the line indicating its $\dot{x}$ and $\dot{y}$ velocities with a length of 0.5~pc corresponding to 1~\kms. Filled symbols mark stars close to the $xy$ plane, and open symbols the ones which are away from it. The black plus sign marks the cluster centre.  }
\label{FigM67Vs}
\end{figure}

\begin{figure}
  \includegraphics[width=\columnwidth]{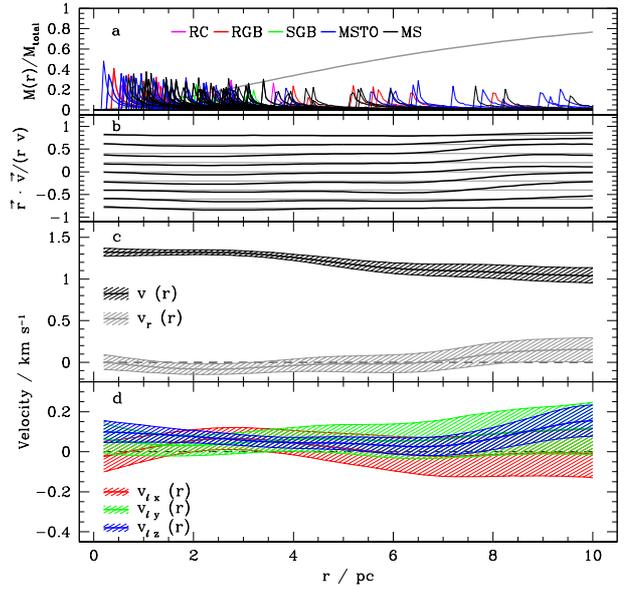}  
  \caption{Large scale motions of stars within M67  and observed by {\sl GALAH$+$}  as a function of distance from the cluster centre. Panel (a) presents distributions of possible positions for each of the observed stars on an arbitrary scale and the mass fraction from the adopted Nuker density profile (in grey). Panel (b) shows the distribution of the cosine of the angle between the vectors $\vec{v}$ and $\vec{r}$. The lines depict the angle at 10, 20, ..., 90\%\ of the distribution: thin grey lines are for an isotropic distribution which has a flat distribution of cosine values between $-1$ and $1$, and thick black lines are for the actual observed distribution. Panel (c) plots the size of the velocity vector ($v(r)$, black) and its projection to the radial direction ($v_r (r)$, grey), while panel (d) shows the three components of the angular velocities ($v_{\ell x}$: red,  $v_{\ell y}$: green, $v_{\ell z}$: blue). Thick lines in panels (c) and (d) are the median values and the shaded regions show the 16\%\ to 84\%\ level spread, as derived from 10,000 realisations of position and velocity of the each of the 193 observed stars. Results in panels b-d are smoothed with a Gaussian with $\sigma = 1$~pc. Zero values in panels (c) and (d) are indicated by a dashed line.}
\label{plot_velocities}
\end{figure}

\begin{figure}
  \includegraphics[width=\columnwidth]{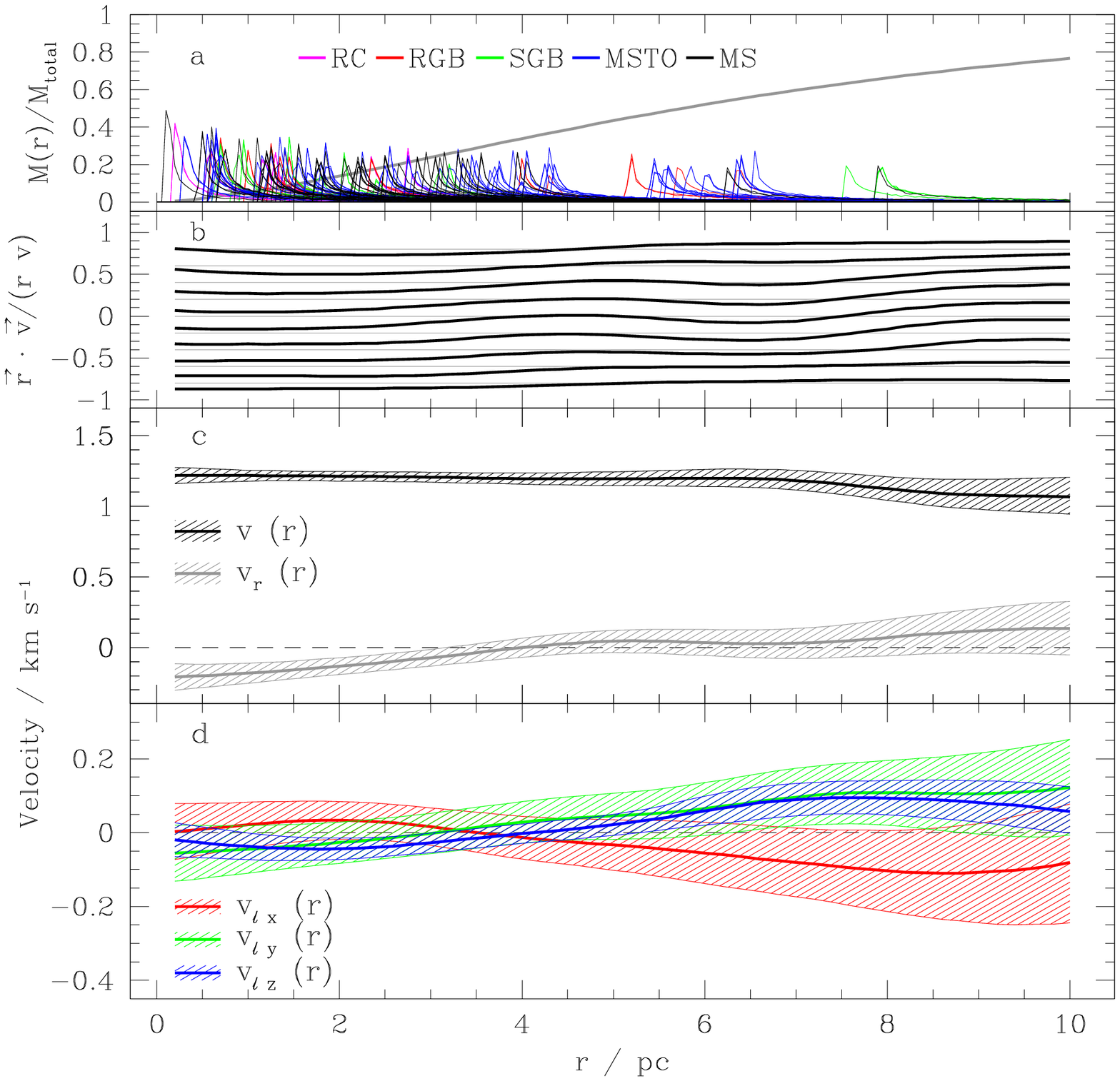}  
  \caption{ As Fig. \ref{plot_velocities}, but for stars observed by {\sl APOGEE} DR16 with their RVs modified, so that they include the effect of gravitational redshift. See text for details. }
\label{plot_velocities_Apogee}
\end{figure}

Figure \ref{FigM67Vs}a plots the $x$ and $y$ position and velocity of each star observed by {\sl GALAH$+$}.  The $z$ position is more uncertain, so we indicate its value by using filled symbols for stars close to the $xy$ plane and open ones for the stars away from it. The latter are stars which have $z^2 > x^2+y^2$ for at least half of their realisations. The $\dot{x}$ and $\dot{y}$ components are shown with a line with a length of 1~pc for each 2~\kms, and the $\dot{z}$ component is indicated by the colour of the symbol. The figure demonstrates that we observed many stars in the central part of the cluster, while those at large distances are mostly offset to negative $y$ values, i.e.\ in the southern direction. This is understandable, as stars with large northern declinations are difficult to observe from the {\sl GALAH}'s southern observing site. Spreads around the cluster centre in the $x$, $y$, and $z$ directions are 1.96, 1.75, and 2.93~pc, respectively. The latter value is largely driven by the assumed Nuker density profile. 
 
Asymmetric spatial distribution of stars observed by {\sl GALAH} may influence the results on their motion with respect to the cluster centre. So we include two independent datasets from the literature. {\sl APOGEE} DR16 \citep{jonsson20} lists observations of 213 stars satisfying the selection criteria discussed above. Their RVs have errors similar to {\sl GALAH}, but they do not include gravitational redshift. So we added this effect (see Sec.~\ref{SecRVs}) using the values of stellar parameters derived by {\sl APOGEE}. \citetalias{brown20} presents astrometric measurements of a complete sample of 808 cluster members which are single stars, but only 62 of these stars have their RVs measured by Gaia and with an average error of 1.3~\kms. So we decided to use a complete sample but with the assumption that all stars have their RVs equal to $\mathrm{RV}_c$. Note that such fixing of one of the components of the velocity vector damps any large scale motions observed within the cluster. Panels \ref{FigM67Vs}b and \ref{FigM67Vs}c plot positions and velocities for the {\sl APOGEE} and Gaia samples. 

Figure \ref{FigM67Vs} presents accurate information on individual velocity vectors and on two out of three spatial coordinates for each of the observed stars. So it is interesting to check if this picture includes some ordered large scale motions within the cluster.  Figure \ref{plot_velocities} presents results for the {\sl GALAH DR3$+$} sample, while { Figure \ref{plot_velocities_Apogee} is its equivalent for RVs measured by {\sl APOGEE}}. Each figure has four panels we discuss next. 

Panel \ref{plot_velocities}{\sl a} illustrates probability distributions of radial positions for each of the stars. The distributions are colour-coded by spectral type of the star, as determined by \citetalias{carrera19} and confirmed by stellar parameters reported in \citetalias{buder20}. Each probability distribution has a sharp peak at a minimal distance $r$ from the cluster centre which is permitted by the $x$ and $y$ coordinates of the star, because the density profile favours small absolute values of $z$. In cases where the parallax value indicates a position in front of or behind the cluster the distribution has a long tail to larger values of $r$. The same panel plots also the mass fraction as a function of radius, as given by the adopted Nuker profile. The function is convex at small radii, but this high density region contains only a small fraction of the total mass. Further out the mass fraction is approximately proportional to the distance, and at  $r>6$~pc it turns to a moderately concave shape, as it includes most of the cluster mass.   

Panel  \ref{plot_velocities}{\sl b} examines the angle between the vectors $\vec{v}$ and $\vec{r}$. It plots the cosine of this angle which has a uniform distribution in the isotropic case. This property is illustrated by uniformly spaced thin grey lines which show the cosine values at 10, 20, ..., 90\%\ of the distribution. However, the thick black lines show the same information, but for the actual observed stars. These lines, and all the other curves in the next panels, are actual values for all realisations of all observed stars, but smoothed with a Gaussian with $\sigma = 1$~pc. Such a choice presents a suitable averaging of individual measurements, yet it preserves the general trends with distance from the cluster centre.

Directions of the velocity vectors in Figure \ref{plot_velocities}b are close to isotropic at all radii. So the stars are generally not in circular orbits. This is demonstrated also by the grey line in Figure \ref{plot_velocities}c, which plots the median value of the radial component of the velocity vector $v_r (r) = \vec{v} \cdot \vec{r} / r$, with the shaded region between 16 and 84 percentiles. Similarly, the black line and its shadowed region in Figure \ref{plot_velocities}c show the size of the velocity vector $v(r)$. 

Subtracting the two curves one can estimate the total mass of the cluster: $M_\mathrm{total} = r \frac{v^2(r)-v_r^2(r)}{G} [\frac{M(r)}{M_\mathrm{total}}]^{-1}$, where the last term is given by the adopted Nuker density profile. For the region with $M(r)/M_\mathrm{total} > 0.5$, corresponding to $r>5.73$~pc, we  get $M_\mathrm{total} = 3300 \pm 100$~M$_\odot$.  There are three reasons why we need to exclude stars close to the cluster centre when estimating the total mass of the cluster: (i) for these stars the uncertainty of the $z$ coordinate increases the fractional uncertainty of their $r$ coordinate, (ii) $M(r) \ll M_\mathrm{total}$ for these stars, so any error in the adopted density profile strongly affects the derived value of the total mass, and (iii) many of the stars in the inner parts of the cluster are close to periastrons of their elliptical orbits, so their velocities are larger than for stars on circular orbits. In particular, the velocity curve $v(r)$ does not decrease towards zero as we approach the cluster centre. Numerical integration of orbits  adopting a stationary Nuker density profile and ignoring any star-star encounters allows us to estimate their typical eccentricities and orbital periods. For the same outer region we obtain eccentricities  $\epsilon \equiv \frac{r_{max}-r_{min}}{r_{max}+r_{min}} = 0.5 \pm 0.2$, with their distribution approximately following a sine-like curve between 0 and 1. So typical stars are on rather elliptical orbits, with $\frac{r_{max}}{r_{min}} \equiv \frac{1+\epsilon}{1-\epsilon} \sim 3$, with the median orbital period of the radial motion  of 88~Myr. Stars which stay closer to the cluster centre have shorter orbital periods, as $M(r) \tilde{\propto}\, r$. The fact that the orbits are not circular is clear already from the directions of the velocity vectors in Figure \ref{FigM67Vs}. The elliptical shape of the stellar orbits may be the reason for the rather large size of the cluster of $\sim 16$~pc  \citep{babusiaux18,gao18}. 

Orbits in the cluster are not symmetrical, but { the inferred radial component of the velocity vector is} small compared to $v(r)$, with absolute values between $0.1$ and $0.2$~\kms. Similarly, Figure \ref{plot_velocities}d { is used to illustrate if stars located away from the cluster core show any net rotation.} By writing the angular momentum $\vec{\ell} (r) = m\, \vec{r} \times \vec{v} = m r \vec{v_\ell}$ we can plot the components of the angular velocity $\vec{v_{\ell}} = (v_{\ell x}, v_{\ell y}, v_{\ell z})$. Note that all these quantities are assumed to depend only on the distance from the cluster centre ($r$), in agreement with the spherically symmetric nature of the Nuker density profile. 

{ Significance of any large scale motions can be judged by comparing results of different surveys which also observe different sets of stars. Figure \ref{plot_velocities_Apogee} does so using RVs measured by the {\sl APOGEE} survey. Results are compatible with {\sl GALAH$+$}. In both cases the significance of non-zero values of $v_r (r)$ and $\vec{v_{\ell}}$ is generally at a one sigma level. Moreover, these results use Gaia eDR3 astrometry which may suffer from spatially correlated systematic errors for objects with angular separations less than one degree, thus relevant for M~67. \citet{vasiliev21} show that this introduces a lower limit on the uncertainty of parallaxes and proper motions at the level of 0.01~mas and 0.025~mas$/$yr, respectively. Parallax errors of M~67 stars quoted by Gaia eDR3 are about twice as large, so effects of systematics are moderate. On the other hand, reported errors on proper motions have an average of 0.023 and 0.015 mas$/$yr for right ascension and declination, respectively. So they may be affected significantly by the possible systematics, which is a consequence of a limited number of scans collected by Gaia over the first 34 months of the mission. 

To address these concerns we tried to quantify contributions of individual types of measurements to the inferred large scale motions in M~67. Appendix \ref{appendix} presents results which are equivalent to Figure \ref{plot_velocities} but obtained by omission of different types of measurements: by applying the cluster average proper motion or RV to all targets, or assuming that their velocities are isotropic.
These tests show that RV measurements by {\sl GALAH$+$} and {\sl APOGEE} hint at a radial expansion of the outer parts of the cluster. Evidence for a possible large-scale rotation is even more uncertain. 

All results on large scale motions have a low statistical significance, largely because of possible systematics affecting the proper motion measurements. This is bound to change with the next data releases of Gaia which will not be affected by the scanning law even for sources at small angular separations. Nevertheless, expansion and rotation of the outer parts of M~67, if confirmed by improved future astrometry, is not unexpected. Despite  its large age, the cluster may not be} completely relaxed. 
This has been suggested before. \citetalias{carrera19} find a number of stars belonging to an extended halo of M~67. They explain their presence by relatively frequent passages of the cluster through the Galactic plane, the last one only $\sim 40$~Myr ago.  \citet{hurley05} use an N-body simulation to show that the cluster lost $\sim 90$\%\ of its initial mass during its evolution. The halo members were not observed by {\sl GALAH}, but here we see a tidal excitation in motions of stars within the cluster. The stars are moving in highly eccentric orbits which may explain the large size of the cluster. These stars are gravitationally bound and given their median period of radial motion of  88~Myr (and its large spread) we note that the last passage through the galactic disc occurred about half of the orbital period ago. As noted by \citetalias{carrera19} the cluster passed the Galactic disc three times in the last 200~Myr. These perturbations keep the cluster in an excited state, so that it did not have enough time yet for a dynamical relaxation. 

\section{Conclusions}
\label{Secconclusions}

In this paper we presented the construction of a new library of observed median spectra observed by the {\sl GALAH$+$} survey, based on parameters of its DR3 data release \citepalias{buder20}. As an example of its use we measured EWs of strong spectral lines of 30 chemical elements across the HR diagram. The observed median spectra are virtually noise-free so that their RVs versus synthetic spectra can be computed over many wavelength intervals which may contain only weak lines. This means that any mismatches in the strength of spectral lines which result from synthetic grid limitations are less important, while moderate displacements of spectral lines which persist over a wide range of median spectra can be used to improve the wavelength solution. The new library and a number of procedure improvements allowed a computation of more accurate RVs of more stars than available before. Altogether we list RVs for 579,653 spectra of 548,056 different stars, with formal velocity uncertainties that are generally smaller than 0.1~\kms. These RVs come in two versions: the values with the gravitational redshift correction are to be used for dynamical studies and when radial velocities of different types of stars are to be compared. The values without the gravitational redshift correction have smaller uncertainties and are useful for studying RV variability or when comparing {\sl GALAH} RVs with other surveys, which generally do not include this correction. As an example, Table \ref{tableRVcomparison} 
compares RVs derived here with the Gaia DR2 and the {\sl APOGEE} DR16 measurements. Median differences are between 9 and 33~m~s$^{-1}$, with an opposite sign of the difference for  dwarfs and giants. The only exception are Gaia DR2 measurements of giant stars where our RVs are $\sim 120$~m~s$^{-1}$ smaller than derived by Gaia. Considering that these surveys do not have their zero points calibrated on each other, we find these results very satisfactory.  

\begin{table}
\caption{Difference in RV between the {\sl GALAH$+$} measurements without the gravitational redshift 
correction presented here and the corresponding values measured by the Gaia DR2 \citep{brown18} and the {\sl APOGEE} DR16 \citep{jonsson20} surveys. Measurements of the latter survey come in 3 variants: average RV (HRV), average RV from observed template technique (HRV2), and average RV from synthetic spectrum template matching technique (HRVs). Only objects with flag\_sp~$=0$ \citepalias{buder20} are considered. The dwarf-giant separation line is defined in eq.\ 1 in \citetalias{zwitter18}. The scatter is half of the difference between the 84.1 and 15.9 percentile levels, calculated after an iterative 3-$\sigma$ clipping.
}
\begin{tabular}{llrcc} 
Survey & Spectral &Spectra      &\multicolumn{2}{c}{$\Delta$~RV /km~s$^{-1}$}\\
            & type       &in common &Median& Scatter \\\hline
Gaia              &dwarfs &93,793 & $+0.009$ & 1.344\\
                     &giants   &110,468& $-0.122$ & 0.761\\
APOGEE(HRV)  &dwarfs & 5,422  & $+0.024$ & 0.376 \\
                      &giants  & 6,520  & $-0.027$  & 0.308 \\
APOGEE(HRV2)&dwarfs & 5,426  & $+0.015 $ & 0.366 \\
                      &giants  & 6,522  & $-0.027$ & 0.302 \\
APOGEE(HRVs) &dwarfs& 5,425  & $+0.033$ & 0.367 \\
                       &giants & 6,521 & $-0.021$  & 0.303 \\\hline
\end{tabular}
\label{tableRVcomparison}
\end{table}

Accurate RVs find their use in detailed studies of Galactic dynamics, which show that our Galactic home is not an ordered equilibrium system, but includes a number of complex oscillations, see a recent demonstration by \citet{antoja20}. It seems that this applies also to motions of stars {\sl within} stellar clusters. In particular, {\sl GALAH$+$} observed 244 members of the open cluster M~67 and determined their RVs. We used \citetalias{brown20} astrometry to construct a probabilistic three dimensional map of positions of these stars within the cluster. By adding RVs we also determined their velocity vectors with respect to the centre of the cluster. The size of the velocity vector is consistent with a total mass of the cluster of $3300 \pm 100$~M$_\odot$. Stars are in elliptical orbits, typical eccentricity is $0.5 \pm 0.2$. 

We realise that the cluster is not at rest, { with some hints of expansion and rotation in its outer parts, though these claims have a low statistical significance. The situation is expected to be clarified when current RV measurements from {\sl GALAH$+$} and {\sl APOGEE} surveys will be combined with future data releases of Gaia that will be free from systematics which may currently affect astrometry of compact sources, such as stars in clusters.} To the best of our knowledge this is the first such kinematic study of an open cluster. It is a witness to the accuracy achievable with { Gaia} astrometry when combined with accurate radial velocities. A similar analysis can be done also for some other stellar clusters and for stellar streams across the Galaxy. 

\section*{Acknowledgments}
We thank the referee for very useful comments on the manuscript. 
Based on data acquired through the Australian Astronomical Observatory, under programmes: A/2013B/13 (The {\sl GALAH} pilot survey); A/2014A/25, A/2015A/19, A2017A/18 (The {\sl GALAH} survey phase 1), A2018 A/18 (Open clusters with HERMES), A2019A/1
(Hierarchical star formation in Ori OB1), A2019A/15 (The {\sl GALAH} survey phase 2), A/2015B/19, A/2016A/22, A/2016B/10, A/2017B/16, A/2018B/15 (The HERMES-TESS program), and
A/2015A/3, A/2015B/1, A/2015B/19, A/2016A/22, A/2016B/12, A/2017A/14, (The HERMES K2-follow-up program). We acknowledge the traditional owners of the land on which the AAT stands, the Gamilaraay people, and pay our respects to elders past and present. TZ, JK, and K\v{C} acknowledge financial support of the Slovenian Research Agency (research core funding No.\ P1-0188) and the European Space Agency (Prodex Experiment Arrangement No.\ C4000127986). Parts of this research were supported by the Australian Research Council Centre of Excellence for All Sky Astrophysics in 3 Dimensions (ASTRO 3D), through project number CE170100013.

This work has made use of data from the European Space Agency (ESA) mission Gaia (http://www.cosmos.esa.int/gaia), processed by the Gaia Data Processing and Analysis Consortium (DPAC, http://www.cosmos.esa.int/web/gaia/dpac/consortium). Funding for the DPAC has been provided by national institutions, in particular the institutions participating in the Gaia Multilateral Agreement.

The following software and programming languages made this research possible: IRAF \citep{tody86,tody93}, SuperMongo (version 2.4.43, developed by Rober Lupton and Patricia Monger); topcat \citep[version4.4,][]{taylor05}; Python and its packages astropy \citep{astropy13}, scipy \citep{jones01}, matplotlib \citep{hunter07}, NumPy \citep{walt11}. This research has made use of the VizieR catalogue access tool, CDS, Strasbourg, France \citep{ochsenbein00}.

\section{Data availability}
\label{Secdataproducts}

This paper has two main data products. A library of median observed stellar spectra can be downloaded from the {\sl GALAH} DR3 website, as explained in \citetalias{buder20}. Similarly, the radial velocity catalogue with and without gravitational redshift corrections can be downloaded as one of the Value Added Catalogues (VACs) from the same website. All observed median spectra and all tables are available also from the first author's homepage\footnote{\url{http://fiz.fmf.uni-lj.si/zwitter/GALAHDR3/}}. Complete versions of Tables  \ref{tableRVs} -- \ref{tableconstantRV} are available online and from the CDS Vizier service.

\noindent \rule{8.5cm}{1pt}

\noindent
$^{1}$Faculty of Mathematics and Physics, University of Ljubljana, Jadranska 19, 1000 Ljubljana, Slovenia\\ 
$^{2}$Research School of Astronomy \& Astrophysics, Australian National University, ACT 2611, Australia\\
$^{3}$Centre of Excellence for Astrophysics in Three Dimensions (ASTRO-3D), Australia\\
$^{4}$Max Planck Institute for Astrophysics, Karl-Schwarzschild-Str. 1, D-85741 Garching, Germany\\
$^{5}$Sydney Institute for Astronomy, School of Physics, A28, The University of Sydney, NSW 2006, Australia\\
$^{6}$Monash Centre for Astrophysics, Monash University, Australia\\
$^{7}$School of Physics and Astronomy, Monash University, Australia\\
$^{8}$Australian Astronomical Optics, Faculty of Science and Engineering, Macquarie University, Macquarie Park, NSW 2113, Australia\\
$^{9}$Macquarie University Research Centre for Astronomy, Astrophysics \& Astrophotonics, Sydney, NSW 2109, Australia\\
$^{10}$Istituto Nazionale di Astrofisica, Osservatorio Astronomico di Padova, vicolo dell'Osservatorio 5, 35122, Padova, Italy\\
$^{11}$Department of Astronomy, Stockholm University, AlbaNova University Centre, SE-106 91 Stockholm, Sweden\\
$^{12}$Max Planck Institute for Astronomy, K\"{o}nigstuhl 17, 69117 Heidelberg, Germany\\
$^{13}$School of Physics, UNSW, Sydney, NSW 2052, Australia\\
$^{14}$Stellar Astrophysics Centre, Department of Physics and Astronomy, Aarhus University, DK-8000, Aarhus C, Denmark\\
$^{15}$Department of Physics and Astronomy, Macquarie University, Sydney, NSW 2109, Australia\\
$^{16}$Research Centre for Astronomy, Astrophysics and Astrophotonics, Macquarie University, Sydney, NSW 2109, Australia\\
$^{17}$Institute for Advanced Study, Princeton, NJ 08540, USA\\
$^{18}$Department of Astrophysical Sciences, Princeton University, Princeton, NJ 08544, USA\\
$^{19}$Observatories of the Carnegie Institution of Washington, Pasadena, CA 91101, USA\\
$^{20}$Lund Observatory, Department of Astronomy and Theoretical Physics, Box 43, SE-221 00 Lund, Sweden\\
$^{21}$Commonwealth Department of Industry, Science, Energy and Resources, 105 Delhi Road, North Ryde, NSW 2113, Australia\\
$^{22}$Centre for Astrophysics, University of Southern Queensland, Toowoomba, QLD 4350, Australia

\appendix
\section{Contributions of individual datasets to stellar motions within M~67}
\label{appendix}

{ 
In the main text we discuss measurements of stellar motions within M~67 as indicated by observations of Gaia and {\sl GALAH$+$} or {\sl APOGEE} surveys. Here we study contributions of individual datasets by selectively omitting some of the astrometric or spectroscopic measurements. 

In Figure \ref{plot_velocities_onlyGalahRV} we omit the proper motion information. All stars are assumed to have their proper motions equal to the scaled value of proper motion of the cluster centre: $\mu_{\alpha,\delta} = (d_c / d) \mu_{\alpha,\delta\, c} $. RVs measured by {\sl GALAH$+$} suggest a radial expansion of the cluster and a rather pronounced rotation in the $(x,z)$ plane. Figure \ref{plot_velocities_onlyApogeeRV} is an equivalent plot, but using stars with RVs measured by the {\sl APOGEE} survey. Note however that these plots still use some astrometric results from Gaia. In particular, parallaxes of individual stars determine the sign of $v_r$ velocities: a star with a RV which is larger than $\mathrm{RV}_c$ indicates an expansion if it lies behind the cluster centre and a contraction if it is located in front of it. Similarly, the value of $v_{\ell y}$ is different for stars which are located further away or closer than the cluster centre.

Figure \ref{plot_velocities_onlyGaiaeDR3} uses astrometric information from Gaia eDR3 for all cluster members, but omits spectroscopically determined RVs. As explained in the main text, RVs determined by Gaia are not sufficiently numerous and accurate for our purpose. So we assumed that all RVs are equal to $\mathrm{RV}_c$. The results are consistent with no internal motions and the error-bars are small because of a large number of stars considered. Note however, that the simplistic assumption of constant RVs for all stars is not realistic. Finally we check what happens if we keep the size of the velocity vector of any star with respect to the cluster centre (as measured by {\sl GALAH$+$} and Gaia) but assume its orientation is isotropic (Fig.\ \ref{plot_velocities_randomV}). If the number of stars observed by {\sl GALAH} were large the average motions would be zero. In reality, small number statistics reflects in velocity deviations within the expected uncertainties. 
 
Figures \ref{plot_velocities_onlyGalahRV}  and  \ref{plot_velocities_onlyApogeeRV} show that both {\sl GALAH$+$} and {\sl APOGEE} RVs favour a radial expansion in the outer parts of the cluster which is much more significant than when using measured proper motions of individual stars (Figures \ref{plot_velocities} and \ref{plot_velocities_Apogee}). The fact that Gaia eDR3 proper motions do not support large scale velocities on the level of 0.1~\kms\ or higher is demonstrated by Figure \ref{plot_velocities_onlyGaiaeDR3}.     
}

\begin{figure}
  \includegraphics[width=\columnwidth]{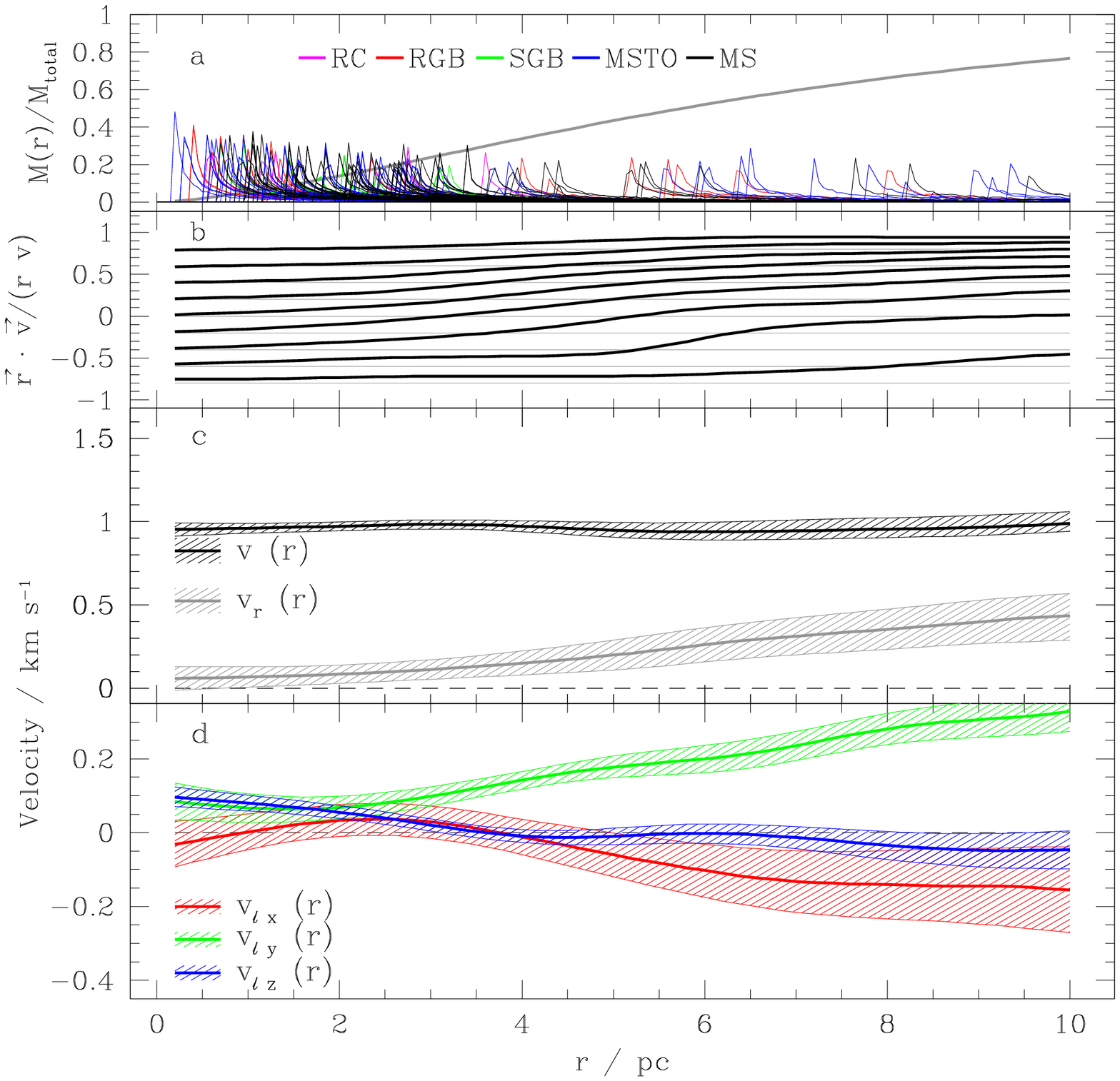}  
  \caption{{ As Fig. \ref{plot_velocities}, but assuming proper motions of all stars are equal to scaled values of the proper motion of the cluster centre: $\mu_{\alpha,\delta} = (d_c / d) \mu_{\alpha,\delta\, c} $. Such a choice emphasises the role of RV measurements by the {\sl GALAH$+$} survey and avoids any systematic errors in proper motion measurements, though astrometry is still used to determine the distance distribution of individual stars.} }
\label{plot_velocities_onlyGalahRV}
\end{figure}

\begin{figure}
  \includegraphics[width=\columnwidth]{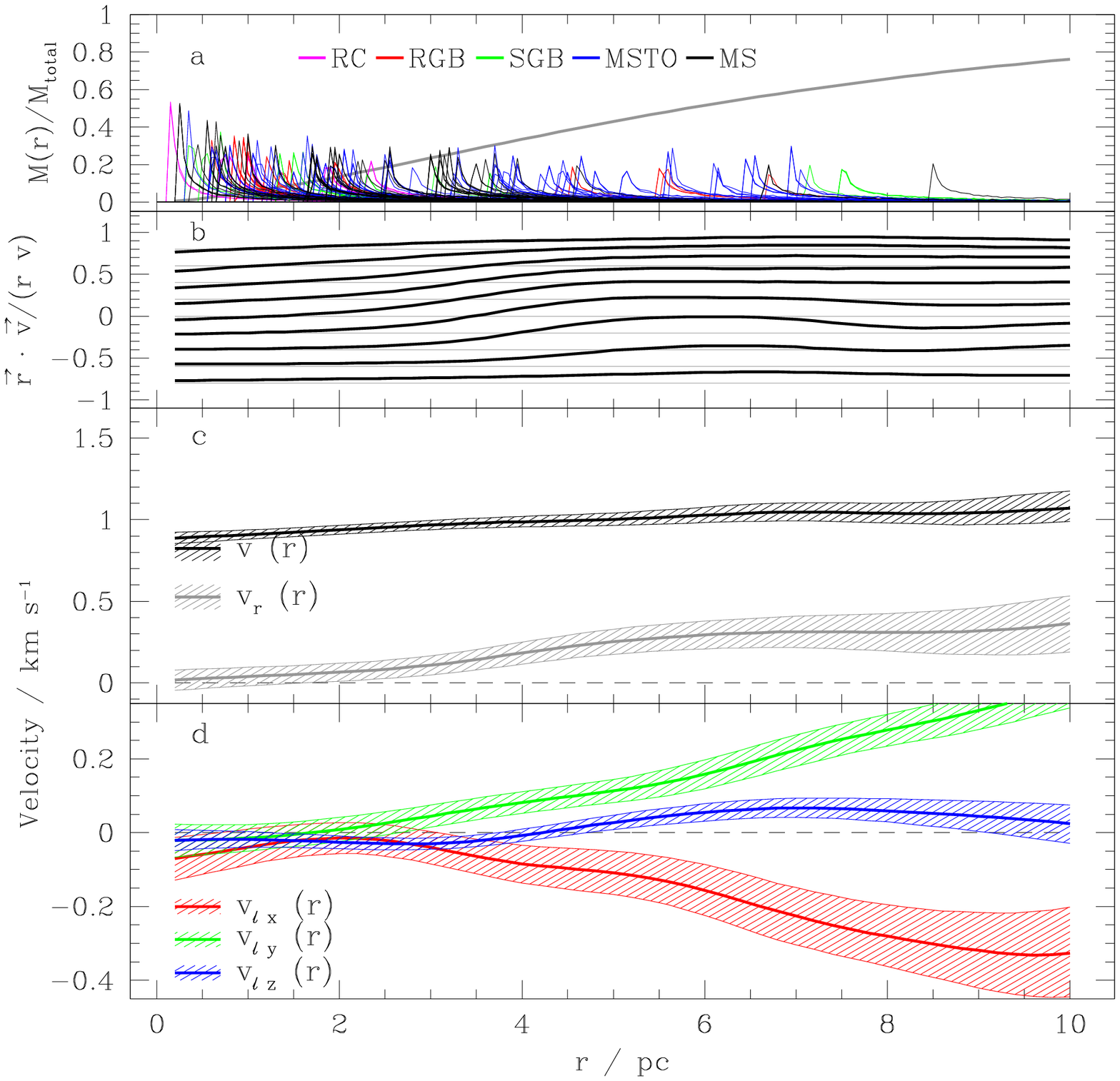}  
  \caption{{ As Fig. \ref{plot_velocities_onlyGalahRV}, but using RVs from the {\sl APOGEE} survey.}}
\label{plot_velocities_onlyApogeeRV}
\end{figure}

\begin{figure}
  \includegraphics[width=\columnwidth]{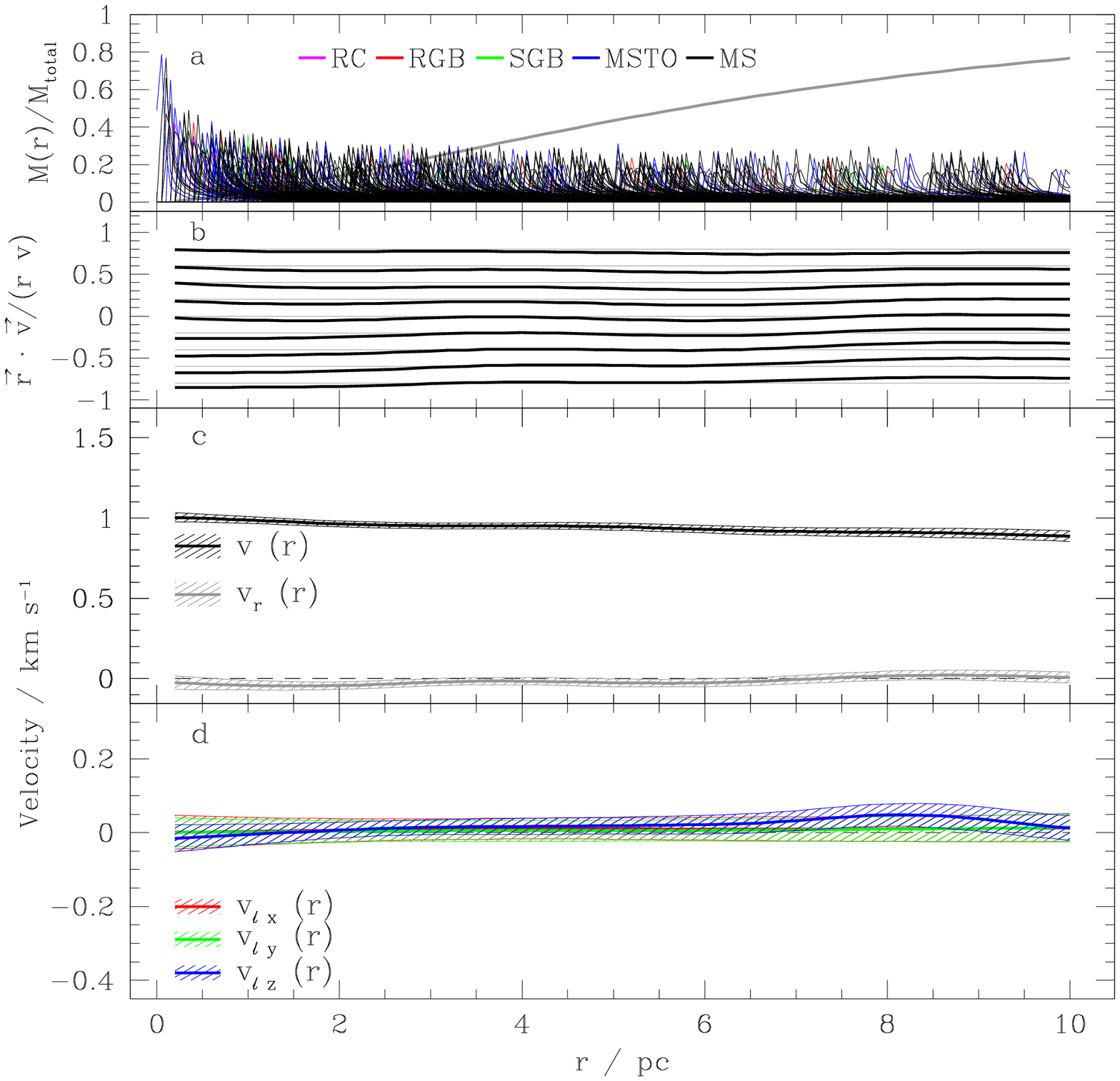}  
  \caption{As Fig. \ref{plot_velocities}, but for all stars in Gaia eDR3. Since only a small fraction of these stars have RVs measured by Gaia and with large errorbars we assumed that all stars have their RVs equal to the cluster velocity ($\mathrm{RV}_c$). This assumption lowers the medians of reported velocities and shrinks their error-bars. }
\label{plot_velocities_onlyGaiaeDR3}
\end{figure}

\begin{figure}
  \includegraphics[width=\columnwidth]{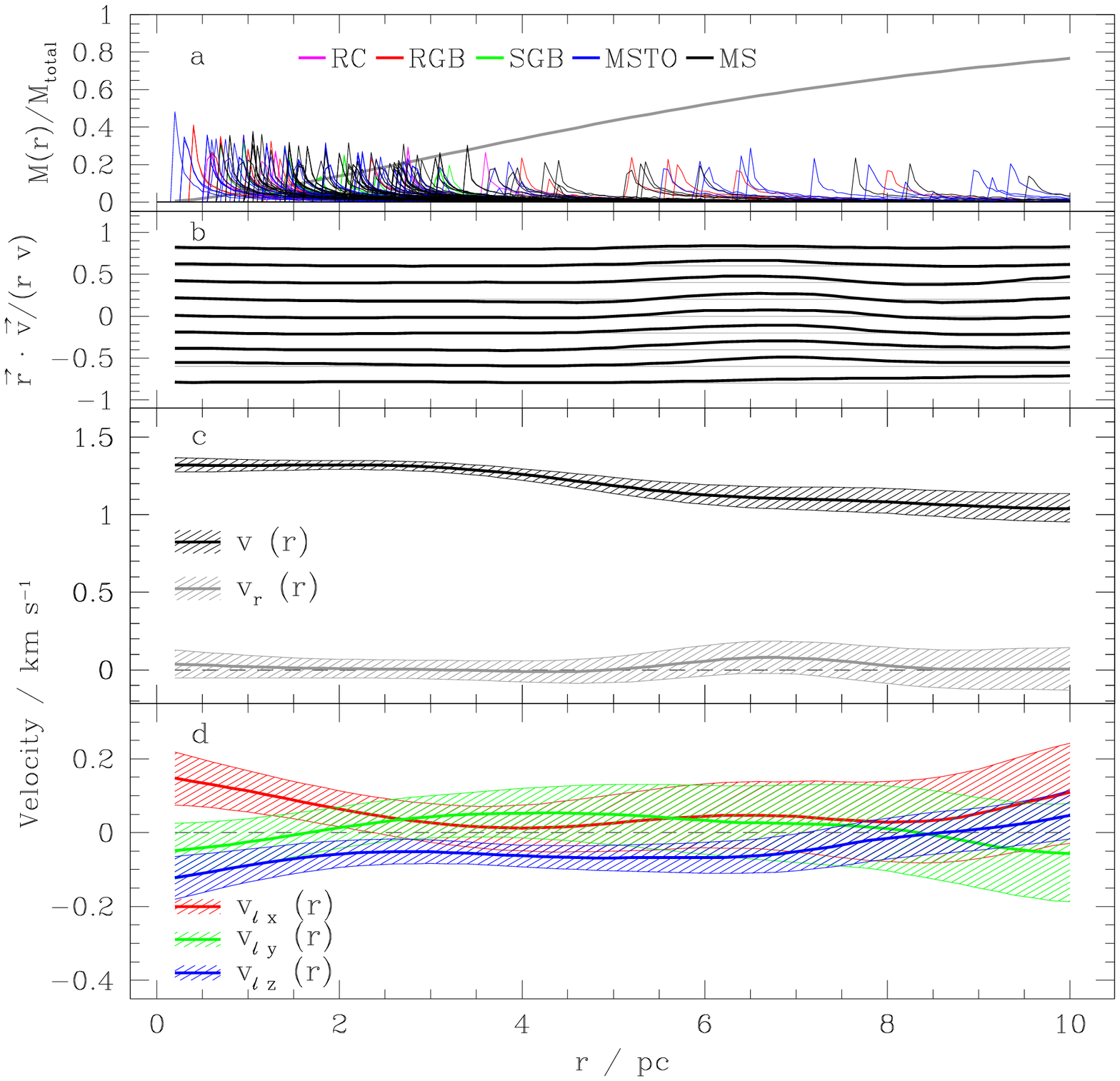}  
  \caption{ { As Fig. \ref{plot_velocities}, but assuming that the velocity vectors of individual stars vs. the cluster centre point to random, isotropic directions, while their size is as measured by {\sl GALAH$+$} and Gaia. Average values of $v_r$ and $\vec{v_\ell}$ are not zero because of a small number statistics which reflects a moderate number of M~67 members observed by {\sl GALAH}.}}  
\label{plot_velocities_randomV}
\end{figure}

\end{document}